\documentclass[twocolumn]{aastex631}
\usepackage{bm}
\usepackage{amsmath}
\usepackage{color}
\usepackage{float}
\renewcommand{\vec}[1]{\mbox{\boldmath$#1$}}
\shorttitle{Tidally forced planetary waves in solar-like stars}
\shortauthors{Horstmann et al.}
\graphicspath{{./}{figures/}}

\begin{document}

\title{Tidally Forced Planetary Waves in the Tachocline of Solar-like Stars}

\author[0000-0001-9892-9309]{Gerrit M. Horstmann}
\affiliation{Institute of Fluid Dynamics, Helmholtz-Zentrum Dresden-Rossendorf, 
Bautzner Landstrasse 400, 01328 Dresden, Germany}

\author[0000-0002-6189-850X]{George Mamatsashvili}
\affiliation{Institute of Fluid Dynamics, Helmholtz-Zentrum Dresden-Rossendorf, 
Bautzner Landstrasse 400, 01328 Dresden, Germany}
\affiliation{Ilia State University, Cholokashvili ave. 5/3, Tbilisi, Georgia}
\affiliation{Abastumani Astrophysical Observatory, Mount Kanobili, Georgia}

\author[0000-0002-2009-3166]{Andr\'{e} Giesecke}
\affiliation{Institute of Fluid Dynamics, Helmholtz-Zentrum Dresden-Rossendorf, 
Bautzner Landstrasse 400, 01328 Dresden, Germany}

\author[0000-0001-5015-5762]{Teimuraz V. Zaqarashvili}
\affiliation{Institute of Physics, IGAM, University of Graz, Universitätsplatz 5, 8010, Graz, Austria}
\affiliation{Ilia State University, Cholokashvili ave. 5/3, Tbilisi, Georgia}
\affiliation{Abastumani Astrophysical Observatory, Mount Kanobili, Georgia}

\author[0000-0002-8770-4080]{Frank Stefani}
\affiliation{Institute of Fluid Dynamics, Helmholtz-Zentrum Dresden-Rossendorf, Bautzner Landstrasse 400, 01328 Dresden, Germany}



\begin{abstract}
Can atmospheric waves in planet-hosting solar-like stars substantially resonate to tidal forcing? Substantially at a level of impacting the space weather or even of being dynamo-relevant? In particular, low-frequency Rossby waves, which have been detected in the solar near-surface
layers, are predestined at responding to sunspot cycle-scale perturbations. In this paper, we seek to address these questions as we formulate a forced wave model for the tachocline layer, which is widely considered as the birthplace of several magnetohydrodynamic planetary waves, i.e., Rossby, inertia-gravity (Poincar\'{e}), Kelvin, Alfv\'{e}n and gravity waves. The tachocline is modeled as a shallow plasma atmosphere with an effective free surface on top that we describe within the Cartesian $\beta$-plane approximation. As a novelty to former studies, we equip the governing equations with a conservative tidal potential and a linear friction law to account for dissipation. We combine the linearized governing equations to one decoupled wave equation, which facilitates an easily approachable analysis. Analytical results are presented and discussed within several interesting free, damped and forced wave limits for both mid-latitude and equatorially trapped waves. For the idealized case of a single tide generating body following a circular orbit, we derive an explicit analytic solution that we apply to our Sun for estimating leading-order responses to Jupiter. Our analysis reveals that Rossby waves resonating to low-frequency perturbations can potentially reach considerable velocity amplitudes in the order of $10^1\,$\textendash$\, 10^2\, {\rm cm}\, {\rm s}^{-1}$, which, however, strongly rely on the yet unknown total dissipation.
	
\end{abstract}

\keywords{Sun: interior – Sun: magnetic fields – Sun: oscillations}


\section{Introduction} \label{sec:intro}
For more than 100 years it has been known that the global weather and climate of our earth is decisively impacted by atmospheric planetary waves, in which large-scale Rossby waves occupy the most prominent role \citep{Pedlosky1987}. But only in the last 20 years strong evidence has accumulated that Rossby waves play similar vital roles in various astrophysical objects, such as planets, e.g., Jupiter \citep{Li2006} or Saturn \citep{Read2009}, accretion discs \citep{Lyra2019} and, most importantly, in the Sun and other solar-like stars, see \cite{Zaqarashvili2021} for an in-depth review. With regard to our Sun, it is known today that Rossby waves are promising candidates for explaining the solar seasons \cite{Dikpati2017,Dikpati2018a}, inducing angular momentum transport \citep{Gizon2020} and impacting or even causing solar activity cycles, starting from Riega-type periodicities \citep{Zaqarashvili2010a} over Schwabe cycle fluctuations \citep{Raphaldini2015,Raphaldini2019} up to long-term modulations in the order of the Gleisberg cycle \citep{Zaqarashvili2018a}. Rossby waves may also serve a crucial role for the solar dynamo \citep[section~5.5]{Zaqarashvili2021} and have even been considered as a key ingredient for dynamo action; an early idea of a self-exciting Rossby wave dynamo dates back to \cite{Gilman1968}.

For a long time, Rossby waves with respect to the Sun were mostly perceived as a theoretical concept and a variety of explanations for their hypothetical emergence were proposed. The breakthrough has only very recently been obtained: different types of Rossby waves have been detected independently by employing different methodologies. First, \cite{McIntosh2017} have observed magnetic Rossby-like waves in the solar atmosphere by tracking coronal bright points from EUV images. Thereupon, \cite{Loptien2018} and \cite{Liang2019} have further observed classic Rossby modes in solar near-surface layers from helioseismic measurements. It is important to note that classic and magnetic Rossby waves are physically very different and have, among other more nuanced dissimilarities, exactly opposite phase velocities and opposite group velocities; see \cite{Dikpati2020} for an excellent introduction to the physics of solar Rossby waves. Both types of waves can therefore play very different roles in the solar dynamics and it is today of great concern to understand the different manifestations of Rossby and other magnetohydrodynamic planetary waves in the Sun \citep{Zaqarashvili2021}.

Today, two of the most important and as yet largely unresolved questions are where and how solar Rossby waves might originate. Although Rossby waves have been observed in the outer solar atmosphere, the shallow tachocline layer is widely believed to be one of the most promising birthplaces of solar planetary waves. \cite{Gilman2000} has shown in a pioneering work that the solar tachocline fluid-layer can be treated, in terms of hydrodynamics, fairly analogously to the lower atmosphere of the Earth, which is why the well-studied geophysical shallow water equations, governing Rossby and other classical atmospheric waves, can be transferred almost one-to-one to the solar tachocline. Since that time, numerous two-dimensional as well as quasi-3D shallow-water models have been employed to study
the global wave instabilities in the tachocline, see, e.g., \cite{Dikpati2001,Schecter2001,Gilman2002,Zaqarashvili2007,Zaqarashvili2009,Zaqarashvili2010,Raphaldini2015,Klimachkov2017,Dikpati2018a}. These studies have identified a number of different possible causes for the development of Rossby waves, including different shear instabilities associated to differential rotation, thermal forcing and nonlinear wave-wave interactions.

One natural creation mechanism, which has gained a special
importance in recent years concerning the solar dynamo, has
been disregarded so far: the possible wave excitation by tidal forcing. It was emphasized in a number of studies \citep{Stefani2016a,Stefani2018,Stefani2019,Stefani2021} that the combined tidal action of the planets Earth, Venus and Jupiter might play a significant role in the synchronization process of the solar dynamo. Although the responding tidal height is only in the order of $1\, {\rm mm}$, energetically equivalent velocities can reach up to $1\, {\rm m}\, {\rm s}^{-1}$ due to the high gravitational acceleration in the tachocline, which could indeed be dynamo-relevant. While in those works the synchronization mechanism was considered to rely only on the entrainment of the $\alpha$-effect caused by the Taylor instability \citep{Weber2015}, the thrilling question has arisen of whether Rossby and other planetary waves could take on a similar facilitating role.
Can Rossby waves possibly intensify the tidal action, or, in other words, may they serve as kind of a resonance ground for tidal excitations? Slow magnetic Rossby waves can have periods in the order of the solar cycle, which is remarkably close to the 11.07 year period visible in the spring-tide envelope curve of Jupiter's, Venus' and Earth's tidal potentials \citep{Okhlopkov2016}. But also in other respects Rossby waves entail excellent premises on which to act on the solar dynamo. First, Rossby waves can have a net kinetic helicity \citep{Dikpati2001,Gilman2002} letting them participate in the $\alpha$-effect. Second, tachoclinic oscillations can further sensitively affect magnetic flux tube instabilities since very small variations of the superadiabacity $\delta$ (stratification of specific entropy) in the order of $\delta \sim 10^{-4}$ to $\delta \sim 10^{-5}$ considerably alter the magnetic storage capacity \citep{Ferriz-Mas1996,Abreu2012,Charbonneau2022}. Motivated by all these auspicious premises, we devote this study to tidally forced magnetic planetary waves and present a first theoretical ``shallow-water'' model, which can account for arbitrary tidal potentials and, as a second novelty, further entails Rayleigh friction permitting the study of damped wave mechanics. 

The paper is organized as follows. In Section \ref{sec:Formulation} we formulate the forced wave model on the local Cartesian $\beta$-plane and rearrange the governing equations into one decoupled wave equation for the latitude velocity, which covers the entire wave physics and makes the analysis easily accessible. In Section \ref{sec:results} we present analytical results for both mid-latitude and equatorially trapped waves. We start to recover the known free wave problems and gradually increase in complexity all the way from different freely damped limits up to the full forced wave problem, which is solved for a single tide generating body. In Section \ref{sec:Sun}, the solutions are finally applied to the particular scenario of tachoclinic waves in our Sun forced by Jupiter to estimate resonant velocity amplitudes.

\section{Formulation of the shallow-water model} \label{sec:Formulation}
\subsection{Basic considerations}
\begin{figure}
\includegraphics[scale=0.95]{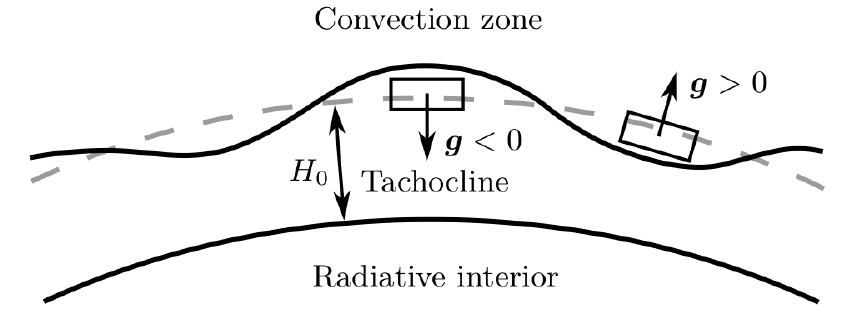}	
	\caption{Schematic drawing of the tachocline modeled as a plasma ocean with the average depth $H_0$ bounded by a rigid bottom and an imaginary free interface at the top. The interface feels an effective gravity $\bm{g}$ arising from the fact that the convection layer, which is adiabatic or even superadiabatic, does not offer any buoyancy resistance to fluid volumes coming from below, whereas fluid volumes within the  subadiabatically stratified tachocline are subject to negative buoyancy.  \label{fig:Tachocline}}
\end{figure}

Shallow water formulations have been widely employed in the last century to describe several types of planetary waves both in the atmosphere and oceans of rotating planets \citep{Longuet-Higgins1964,Pedlosky1987}\textemdash with Rossby waves perhaps being the most prominent member. In its simplest view, an ocean can be intuitively modeled as a one-layer system of fluid sandwiched between a flat rigid bottom and a freely movable surface placed on top. If we focus on large-scale planetary waves, this layer can be considered as ‘shallow’ in the sense that characteristic wave lengths are much larger than the average water depths. This has the consequence that the flow is of quasi-two-dimensional nature, i.e., the velocities do not alter significantly with altitude and the change of momentum caused by vertical velocities and vertical Coriolis force components can be neglected in the momentum conservation law. Such shallow water approaches are very intelligible and simplify the mathematical analysis enormously. Therefore, it is all the more gratifying that planetary waves evolving in tachoclines of solar-like stars can be described in very similar ways. It was pointed out in the pioneering paper by \cite{Gilman2000} that the geophysical shallow water equations can be transferred almost one-to-one to the solar tachocline. In a manner of speaking, the tachocline can be treated as a sort of plasma ocean as schematized in Figure \ref{fig:Tachocline}. 
The tachocline layer with the depth $H_0$, which itself consists of a radiative part and an overshoot layer (not shown here), is the transition region between the radiative interior and the outer convection zone. The convection zone is of adiabatic or even superadiabatic nature, whereas the stably stratified tachocline is subadiabatic. Hence, warmer fluid volumes entering the convection zone tend to rise (or at least do not experience any
buoyancy resistance in the adiabatic case), while warmer fluid volumes tend to sink in the tachocline. As a result of this behavior, the transition region between the tachocline and the convection zone can be replaced by an imaginary interface, which experiences some effective gravity in the same way as ocean-atmosphere interfaces are exposed to standard gravity. The effective gravity force is proportional to the fractional
difference between the actual and adiabatic temperature
gradients $\sim |\nabla - \nabla_{\rm ad}|$, taking values of $10^{-4} - 10^{-6}$ in the upper overshoot part of the tachocline and up to $10^{-1}$ in the lower radiative sublayer \citep{Dikpati2001}. These gradients yield very different effective gravity constants of $g = 0.05 - 5 \,{\rm cm}\, {\rm s}^{-2}$ (overshoot part) and $g = 500 - 1.5\cdot 10^4 \,{\rm cm}\, {\rm s}^{-2}$ (radiative sublayer) \citep{Schecter2001}.

The shallow water equations of the plasma ocean model can be applied separately to both parts of the tachocline (but not both at the same time) such that the natural frequencies of planetary waves vary by orders of magnitude among the sublayers. The density is approximated to be constant and the lower interface to be stationary since the higher dense radiative interior appears much more rigid than the tachocline layer above. Note that there are enhanced two-layer and even continuous descriptions \citep{Hunter2015,Petrosyan2020,Fedotova2021}, the here adopted one-layer system, however, is well established and contains the essential wave dynamics, particularly with regard to Rossby waves, which allowed to make significant contributions to various astrophysical settings \citep{Zaqarashvili2021}.   

\subsection{Governing equations \label{subsec:Governing equations}}
We consider a thin layer of an ideally conductive, incompressible and inviscid fluid on a sphere within the rotating frame of reference. The angular velocity $\bm{\Omega}_0$ as well as the density $\rho$ are set to be constant. The layer of height $H_0$ is further traversed by a magnetic field $\bm{B}$,
which is nearly horizontal in the solar tachocline. As the key new feature, we also take into account external tidal forces, which can be uniquely expressed through the gradient of a conservative tidal potential $V$. Since the inviscid models lack any dissipation source, the inclusion of an external force gives rise to the problem of singularities appearing under resonance conditions. In geophysical flows it is common to include linear friction terms, usually Rayleigh friction and/or Newtonian cooling (see, e.g., \cite{Gill1980,Mofjeld1981,Yamagata1985,Wu2001}), which are indicative in the framework of linear waves. Here, we include Rayleigh's friction law, stating that, in the leading order, the effective drag is proportional to the flow velocity and some \textit{a priori} unknown damping constant $\lambda$. Although viscous and thermal dissipation in the tachocline is highly nonlinear on smaller scales, the linear friction law can nevertheless be an appropriate approximation for many long-period and large-scale planetary waves\textemdash at least when the wave amplitudes are sufficiently small. 

By equipping the classical magnetohydrodynamic shallow water equations proposed by \cite{Gilman2000} with tidal and friction forces, our governing equations can be expressed in the following way:
\begin{align}
	&\frac{\partial \bm{u}}{\partial t} + (\bm{u}\cdot\bm{\nabla})\bm{u} +2\bm{\Omega}_0 \times \bm{u} +\lambda \bm{u} \nonumber \\ 
	&=-g\bm{\nabla}H 
	 + \frac{1}{4\pi \rho}(\bm{B}\cdot\bm{\nabla})\bm{B} - \bm{\nabla}V , \label{eq:Mom}\\
	&\frac{\partial \bm{B}}{\partial t} + (\bm{u}\cdot\bm{\nabla})\bm{B} = (\bm{B}\cdot\bm{\nabla})\bm{u}, \label{eq:Ind}\\
	&\frac{\partial H}{\partial t} + \bm{\nabla}\cdot(H\bm{u}) = 0, \label{eq:Cont}\\
	&\bm{\nabla}\cdot (H\bm{B}) = 0,\label{eq:DivFree}
\end{align}  
where $\bm{u}$, $\bm{\Omega}_0$ and $\bm{B}$ are the horizontal velocity, angular velocity of the rotating star and magnetic field, $H$ is the total layer height, $\rho$ is the fluid density, $g$ is effective gravitational acceleration and $\bm{\nabla}$ denotes here the purely horizontal gradient. Equation (\ref{eq:Mom}) is the Euler equation including the Coriolis force, the Lorentz force, gravity and Rayleigh friction. Equation (\ref{eq:Ind}) is the induction equation of the magnetic field in the limit of ideally conducting fluids (high magnetic Reynolds number limit) and (\ref{eq:Cont}) is the shallow water version of the continuity equation. Equation (\ref{eq:DivFree}) finally ensures that the magnetic field is divergence-free on the condition that $\bm{B}$ remains parallel to the upper free surface.  
\subsection{$\beta$-plane approximation} 
\begin{figure}
\begin{centering}
\includegraphics[scale=1.1]{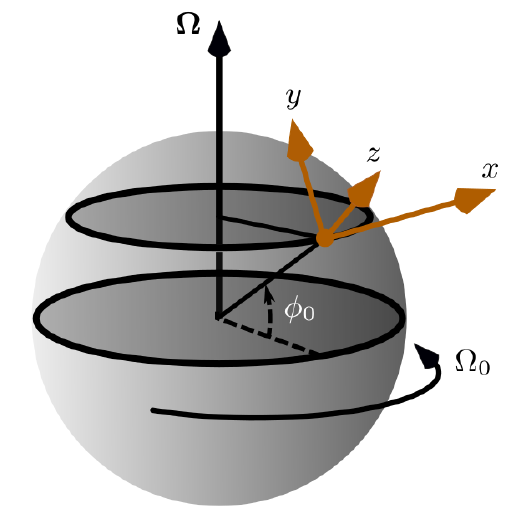}
\end{centering}	
	\caption{Local Cartesian coordinate system defined at a fixed latitude $\phi_0$ in the non-inertial frame of reference on the tachocline surface of a sphere constantly rotating with angular velocity $\Omega_0$. The $x$-axis points eastwards (longitude), the $y$-axis northwards (latitude) and the $z$-axis radially outwards from center (altitude).  \label{fig:BetaPlane}}
\end{figure}

The inclusion of tidal forcing drastically enriches the physical complexity. For this study, we are mainly interested in finding analytical solutions rather than conducting simulations to understand the incoming physics in a fundamental way. Therefore, in order to keep the analysis as simple as possible, we are going to study the problem in a simpler Cartesian coordinate system within the framework of the so-called $\beta$-plane approximation, see Figure \ref{fig:BetaPlane}. The $\beta$-plane can properly
describe non-equatorial planetary waves if the wave length is sufficiently smaller than the size of the sphere, and is perfectly valid for waves trapped at the equator, where they have exactly the same dispersion relation as in the analogues spherical coordinate system \citep{Pedlosky1987}.  The local Cartesian coordinate system is fixed at the latitude $\phi_0$, at which we can evaluate the Coriolis force
\begin{align}
	&2\bm{\Omega}\times \bm{u} = 2\Omega_0 \begin{pmatrix} 0 \\ \cos \phi_0 \\ \sin \phi_0 \end{pmatrix} \times \begin{pmatrix} u \\ v \\ w \end{pmatrix} \nonumber \\ = \ &2\Omega_0 \begin{pmatrix} v \sin \phi_0 - w \cos \phi_0 \\ -u \sin \phi_0 \\ u  \cos\phi_0 \end{pmatrix} \approx f \begin{pmatrix} v \\ -u \\ 0 \end{pmatrix},
\end{align}
where $u,v,w$ are the longitude, latitude and altitude velocity components and $f = 2\Omega_0 \sin(\phi_0)$ is the Coriolis parameter. The shallow water approximation was applied at the last step, demanding that both the altitude velocity and Coriolis force components are negligibly small, $w, u \cos(\phi_0 ) \approx 0$. It was the great pioneering contribution of Carl-Gustaf Arvid \cite{Rossby1939} to locally expand $f$ on the Cartesian plane for small $\phi_0$ variations as
\begin{align}
	f \approx f_0 + \beta y, \label{eq:f}
\end{align}
with
\begin{align}
	\beta = \frac{{\rm d}f}{{\rm d}y} = \frac{2\Omega_0}{R_0}\cos(\phi_0),
\end{align}
today referred to as the Rossby parameter.
Keeping only the first order $f=f_0 = 2\Omega_0 \sin(\phi_0)$ leads to the $f$-plane approximation, involving different types of magneto-gravity waves (also magneto-Poincar\'{e} waves) and  purely magnetohydrodynamic Alfv\'{e}n waves \citep{Schecter2001}, or also magnetostrophic waves evolving under the presence of vertical magnetic fields \citep{Klimachkov2016}. Accordingly, if one further considers the leading order latitudinal variation of the Coriolis force (second term), we arrive at the so called $\beta$-plane approximation comprising low-frequency magneto-Rossby waves in addition to the $f$-plane waves.

As the last step of approximation, we restrict our analysis to the linear problem, which allows us to derive analytical solutions at the little expense of being limited to small-amplitude waves. By projecting the governing equations (\ref{eq:Mom}) - (\ref{eq:DivFree}) onto the $\beta$-plane and linearizing, we get the equation system 
\begin{eqnarray}
	\frac{\partial u}{\partial t} = - g\frac{\partial \eta}{\partial x} + fv + \frac{B_0}{4\pi \rho}\frac{\partial b_x}{\partial x} - \frac{\partial V}{\partial x} - \lambda u, \label{eq:TD1}\\
	\frac{\partial v}{\partial t} = - g\frac{\partial \eta}{\partial y} -fu + \frac{B_0}{4\pi \rho}\frac{\partial b_y}{\partial x} - \frac{\partial V}{\partial y} - \lambda v, \label{eq:TD2} \\
	\frac{\partial b_x}{\partial t} = B_0 \frac{\partial u}{\partial x}, \ \ \frac{\partial b_y}{\partial t} = B_0 \frac{\partial v}{\partial x}, \label{eq:TD3}\\
	\frac{\partial \eta}{\partial t} + H_0 \left(\frac{\partial u}{\partial x} + \frac{\partial v}{\partial y}\right) = 0, \label{eq:TD4}
\end{eqnarray}
for the individual vector components. Here, $u,v,b_x$ and $b_y$ are the perturbational field quantities and $\eta = H - H_0$ denotes the perturbed layer thickness ($\equiv$ wave amplitude). The equations were perturbed with respect to a uniform toroidal magnetic field $B_0$, which is dominant in the solar tachocline, although generally latitude-dependent. The set of equations is analogues to the free-wave problem analyzed by \cite{Zaqarashvili2007}, with the sole difference that two additional terms appear in the momentum equations (\ref{eq:TD1}) and (\ref{eq:TD2}), addressing tidal forcing and damping. 
\subsection{Decoupled wave equation} 
Although the linearized $\beta$-plane approximation is by far the most accessible description of magneto-planetary waves, the associated wave physics is still very rich. In order to simplify the following analyses and to unify the free wave physics discussed in preceding studies, it is of interest to find a generalized wave equation, which allows the straightforward calculation of dispersion relations in all limiting cases of interest. We found that a fully decoupled wave equation can be derived only for the latitude velocity $v$, but not for all other field variables $u, \eta, b_x$ or $b_y$. After several algebraic transformations, see Appendix \ref{sec:WaveDer}, the set of Equations (\ref{eq:TD1}) - (\ref{eq:TD4}) can be rearranged into the following wave equation:
\begin{widetext}
\begin{equation}
	\square^2_{v_A}v - C_0^2 \square_{v_A} \Delta v + f^2 \frac{\partial^2 v}{\partial t^2} - C_0^2 \beta \frac{\partial}{\partial x}\frac{\partial v}{\partial t} 
	+ \frac{\partial}{\partial t}\frac{\partial}{\partial y}\square_{v_A}V - f\frac{\partial}{\partial x}\frac{\partial^2 V}{\partial t^2} + \lambda\frac{\partial}{\partial y}\frac{\partial^2 V}{\partial t^2} 
	+2 \lambda \frac{\partial}{\partial t}\square_{v_A}v - \lambda C_0^2 \Delta \frac{\partial v}{\partial t} + \lambda^2 \frac{\partial^2 v}{\partial t^2}    = 0, \label{eq:Wave}
\end{equation} 
\end{widetext} 
where $C_0 = \sqrt{gH_0}$ and $v_A = B_0 / \sqrt{4\pi \rho}$ are gravity wave and Alfv\'{e}n velocities, and $\square_{v_A} := \partial_t^2 - v_A^2 \partial_x^2$ denotes the d'Alembert operator with respect to Alfv\'{e}n waves. Equation (\ref{eq:Wave}) serves as the only governing equation throughout the rest of the paper.  
\section{Analytical results}\label{sec:results}
We shall now deduce and discuss tachoclinic wave dynamics within several different wave limits, comprising unbounded ($f \approx f_0$) and equatorially trapped ($f \approx \beta y$), hydrodynamic ($v_A = 0$) and magnetic, inviscid ($\lambda = 0$) and damped, as well as free ($V=0$) and forced waves. First, we reiterate some  basic result on the free wave problem in order to then increase progressively in complexity towards the full tidally forced wave problem. 
\subsection{Free wave dynamics} 
When neglecting the tidal potential, $V = 0$, and damping, $\lambda =0$, Equation (\ref{eq:Wave}) drastically simplifies to the free wave equation 
\begin{equation}
\square^2_{v_A}v - C_0^2 \square_{v_A} \Delta v + f^2 \frac{\partial^2 v}{\partial t^2} - C_0^2 \beta \frac{\partial}{\partial x}\frac{\partial v}{\partial t} 
= 0. \label{eq:Free}
\end{equation}
If one further considers non-magnetic waves $v_A = 0$, (\ref{eq:Free}) transforms into
\begin{equation}
	\frac{\partial^3 v}{\partial t^3}  + f^2	\frac{\partial v}{\partial t} - C_0^2 \frac{\partial }{\partial t}\Delta v - C_0^2 \beta 	\frac{\partial v}{\partial x}  = 0,
\end{equation}
which is the classic planetary wave equation describing Rossby, Poincar\'{e}, Kelvin and gravity waves \citep{Pedlosky1987}. In the following, we want to shortly recapitulate the wave physics captured in Equation (\ref{eq:Free}) within two different limits. First, we focus on non-equatorial waves in the latitude range $30^{\circ} \lesssim \phi_0 \lesssim 60^{\circ}$, where we readily find $f_0 \gg \beta y$ allowing to set $f \approx f_0$. In the vicinity of the equator, we observe $f_0 \ll \beta y$ in contrast, such that one must keep the $y$-dependent term of the Coriolis parameter $f \approx \beta y$, which results in a nonlinear wave equation coming along with fundamentally different wave dynamics.
\subsubsection{Non-equatorial waves}  
\label{sec:FreeNonEq}
Applying $f = f_0$ to Equation (\ref{eq:Free}) and inserting a simple 
Fourier ansatz of the form
\begin{equation}
v = v_0\exp(ik_x x + ik_y y - i\omega t),	
\end{equation}
where $v_0$ is an arbitrary wave amplitude, $k_x$ and $k_y$ are the longitudinal and latitudinal wave numbers and $\omega$ is the angular frequency, we can straightforwardly deduce the following fourth-order dispersion relation
\begin{align}
	\omega^4 - \left[2k_x^2 v_A^2 + f_0^2 + C_0^2(k_x^2 + k_y^2)\right]\omega^2 - C_0^2k_x \beta \omega \nonumber \\
	+k_x^2 v_A^2 \left[k_x^2 v_A^2 + C_0^2(k_x^2 + k_y^2)\right] = 0. \label{eq:DispFree}
\end{align}
This dispersion relation was first derived and discussed by \cite{Zaqarashvili2007}. Let us first consider the limit $\beta = 0$ corresponding to the $f$-plane approximation. In this case, only symmetric frequency polynomials $\omega^4 , \omega^2 , \omega^0$ remain in Equation (\ref{eq:DispFree}), allowing us to rearrange it into the explicit form 
\begin{align}
	&\omega_{\pm}^2 = k_x^2 v_A^2 + \frac{f_0^2}{2} + \frac{C_0^2}{2}\Bigg[ \left(k_x^2 + k_y^2\right) \nonumber \\
	 &\pm \sqrt{\frac{f_0^4}{C_0^4} + \frac{f_0^2}{C_0^2}\left[\frac{4k_x^2 v_A^2}{C_0^2} + 2(k_x^2 + k_y^2)\right] + \left(k_x^2 + k_y^2\right)^2}\Bigg].
\end{align}
This dispersion relation was analyzed in depth by \cite{Schecter2001}, who referred the fast $\omega_+$ branch to magnetogravity waves and the slow $\omega_-$ modes to Alfv\'{e}n waves.
The $\omega_+$ branch can be approximated as 
\begin{align}
	\omega^2 \approx f_0^2 + C_0^2(k_x^2 + k_y^2) + 2k_x^2 v_A^2
	, \label{Inert}
\end{align}
showing that magnetogravity waves are magnetically modified Poincar\'{e} waves, which appear on time scales less than $2\pi/f_0$. The Alfv\'{e}n branch is intrinsically connected to the magnetic field and disappears for $v_A \longrightarrow 0$. Alfv\'{e}n waves can become arbitrarily slow at large length scales ($k_x , k_y \longrightarrow 0$).

Keeping the next order in the expansion (\ref{eq:f}) ($\beta \neq 0$) gives rise to the emergence of Rossby waves. They appear in the low frequency limit $\omega \ll f_0$ of Equation (\ref{eq:DispFree}), where the dispersion relation reads
\begin{align}
	\omega^2 + \frac{R_R^2\beta k_x}{1 + R_R^2 (k_x^2 + k_y^2)}\omega - \frac{R_R^2 v_A^2 k_x^2 (k_x^2 + k_y^2)}{1 + R_R^2 (k_x^2 + k_y^2)} = 0, \label{eq:DispSlow}
\end{align}
here expressed in terms of the Rossby radius $R_R = C_0/f_0$. The Rossby radius describes the length scale at which Coriolis force-driven inertial waves become equally significant as buoyancy-driven gravity waves in the spatiotemporal evolution of linear disturbances. Interestingly, $R_R$ strongly varies throughout the tachocline. One can estimate $R_R \sim 10^4 - 10^6\, {\rm km}$ in the overshoot layer but larger values $R_R \sim 10^7 - 10^8\, {\rm km}$ in the radiative part  of the tachocline at the latitude $\phi_0 = 30^{\circ}$. The tachocline radius is approximately $R_0 \approx 5\cdot 10^5\, {\rm km}$. In the radiative sublayer, we can safely assume $R_R \gg R_0 \gtrsim 1/k_x , 1/k_y$, allowing us to simplify Equation (\ref{eq:DispSlow}) into the pure magneto-Rossby wave dispersion relation 
\begin{align}
	\omega^2 + \frac{\beta k_x}{k_x^2 + k_y^2}\omega - v_A^2 k_x^2 = 0. \label{eq:Rossby}
\end{align}
It contains a high frequency solution
\begin{align}
	\omega \approx -\frac{k_x \beta}{k_x^2 + k_y^2} \label{eq:Classic}
\end{align}
as well as a low frequency branch
\begin{align}
	\omega \approx \frac{k_x v_A^2 (k_x^2 + k_y^2)}{\beta} \label{eq:MagnetoRossby}
\end{align}
in the limit of large length scales. Equation (\ref{eq:Classic}) is the very classic dispersion relation of hydrodynamic Rossby waves. The minus sign reveals a retrograde propagation relative to the rotation of the reference frame. Equation (\ref{eq:MagnetoRossby}) describes the class of magneto-Rossby waves arising only by the effect of horizontal magnetic fields $v_A \neq 0$. These wave modes are quite similar to classic Rossby waves, but move in the prograde direction and have far slower eigen-frequencies, see \cite{Zaqarashvili2007} for more details. A very illustrative description of both classic and magneto-Rossby waves is further given in \cite{Dikpati2020}.
\subsubsection{Equatorial waves}
In the vicinity of the equator, the Coriolis force approximates to $f \approx \beta y$ in the leading order. This yields the wave equation  
\begin{equation}
	\square^2_{v_A}v - C_0^2 \square_{v_A} \Delta v + \beta^2 y^2 \frac{\partial^2 v}{\partial t^2} - C_0^2 \beta \frac{\partial}{\partial x}\frac{\partial v}{\partial t} 
	= 0 \label{eq:FreeEquator}
\end{equation}
as the counterpart to Equation (\ref{eq:Free}). Equation (\ref{eq:FreeEquator}) is nonlinear in $y$ therefore it is convenient to first perform a Fourier analysis for the $x$-part only 
\begin{align}
v = v_y(y)\exp(ik_x x - i\omega t)
\end{align}
yielding 
\begin{equation}
	\frac{{\rm d}^2 v_y}{{\rm d} y^2} + \left[\frac{\omega^2 - k_x^2 (C_0^2 + v_A^2)}{C_0^2} - \frac{k_x \beta \omega}{\omega^2 - k_x^2 v_A^2} - \mu^2 y^2 \right] v_y = 0 \label{eq:Cylindrical}
\end{equation}
for the latitude-dependent velocity part $v_y$. The parameter $\mu$, which is a reciprocal measure for the wave's equatorial expansion, is given as
\begin{align}
	\mu = \frac{\beta \omega}{C_0 \sqrt{\omega^2 - k_x^2 v_A^2}}.
\end{align}  
The differential Equation (\ref{eq:Cylindrical}) was first derived by \cite{Zaqarashvili2018a} and can be identified as the classic equation of the quantum harmonic oscillator. It has bounded solutions of the form
\begin{align}
	v_y = v_0 \exp\left(-\frac{|\mu|y^2}{2}\right)H_n (\sqrt{|\mu|}y),
\end{align}
if and only if  
\begin{align}
	\frac{\omega^2 - k_x^2(C_0^2 +v_A^2)}{C_0^2} - \frac{k_x \beta \omega}{\omega^2 - k_x^2 v_A^2} = |\mu|(2n+1). \label{eq:Contraint}
\end{align}
Here, $v_0$ is some arbitrary amplitude and $H_n$ the Hermite polynomial of order $n$. The solutions are oscillatory inside the latitude interval
\begin{equation}
	y < \left| \sqrt{\frac{2n+1}{|\mu|}}\right| \label{eq:LatExt}
\end{equation}
and exponentially tend to zero outside. Although we have not incorporated any boundary conditions, we arrive at a natural latitudinal boundary condition capturing the waves around the equator. The integers $n$ can be identified as discrete latitudinal wave numbers specifying the number of vortices within the equatorial band defined by Equation (\ref{eq:LatExt}).

The constraint (\ref{eq:Contraint}) defines the dispersion that can be expressed in the following form
\begin{align}
	 (\omega^2 - k_x^2 v_A^2)(\omega^2 - k_x^2(C_0^2 + v_A^2)) - k_x \beta C_0^2 \omega \nonumber \\
	  =  \beta C_0|\omega|\sqrt{\omega^2 - k_x^2 v_A^2}(2n+1). \label{eq:DispEq}
\end{align} 
Neglecting the magnetic field, $v_A = 0$, we arrive at the classic geophysical dispersion relation
\begin{align}
	\omega^2 -  k_x^2C_0^2  - \frac{k_x \beta C_0^2}{\omega} 
	=  \beta C_0(2n+1) \label{eq:DispCl}
\end{align}
\citep{Matsuno1966} comprising gravity-inertia and Rossby waves. The essential novelty of Equation (\ref{eq:DispEq}) compared to the hydrodynamic case (\ref{eq:DispCl}) is that the toroidal field creates low frequency cut-off areas at $\omega = \pm k_x v_A$, suppressing the low-frequency Rossby modes from the hydrodynamic solution. This behavior can, however, change drastically when considering inhomogeneous toroidal magnetic field profiles $B_x \sim B_0 y /R$, giving rise to super-slow magneto-Rossby waves, which can reach periods up to the order of the $100\, {\rm yr}$ Gleissberg cycle \citep{Zaqarashvili2018a}.

The dispersion relation (\ref{eq:DispEq}) can only be solved numerically owing to the square root on the right-hand side.
There are, however, some approximated solutions for certain interesting limit. Restricting to high frequencies $\omega \gg k_x v_A$, Equation (\ref{eq:DispEq}) approximates to
\begin{align}
	\omega^2 \approx \beta C_0 (2n+1) + k_x^2 C_0^2 + 2k_x^2 v_A^2,
\end{align}
describing magneto-inertia-gravity waves, which propagate always faster than the hydrodynamic counterparts. By eliminating the high frequency branches from Equation (\ref{eq:DispEq}), one finds the low-frequency dispersion relation describing Rossby waves
\begin{align}
	k_x^2 \omega^2 + \beta k_x \omega -k_x^4 v_A^2 = - \frac{\beta}{C_0}|\omega|\sqrt{\omega^2 - k_x^2 v_A^2}(2n+1) \label{eq:DispEqRos}
\end{align}
which also is not amenable to analytical solution. We can only find explicit solutions in the limit of weak magnetic fields $v_A \ll 1$, where Equation (\ref{eq:DispEqRos}) reduces to 
\begin{align}
	\omega^2 + \frac{2C_0 \beta k_x\omega}{C_0 k_x^2 + \beta (2n+1)} -\frac{ v_A^2k_x^2 [2C_0 k_x^2 + \beta (2n+1)]}{2[C_0 k_x^2 + \beta (2n+1)]} =0,
\end{align}
which involves again a high-frequency  solution
\begin{align}
	\omega \approx - \frac{C_0 \beta k_x}{C_0 k_x^2 + \beta (2n+1)} \label{eq:DispHD}
\end{align}
and a low-frequency solution 
\begin{align}
	\omega \approx k_x v_A^2 \left(\frac{k_x^2}{\beta} + \frac{1}{2}\frac{2n+1}{C_0}\right). \label{eq:DispEqSlow}
\end{align}
Equation (\ref{eq:DispHD}) is the classic dispersion relation of hydrodynamic, equatorially trapped Rossby waves. Similarly as for the non-equatorial waves, Equation (\ref{eq:DispEqSlow}) describes prograde magneto-Rossby waves and reduces exactly to  Equation (\ref{eq:MagnetoRossby}) in the limit of purely longitudinally propagating waves ($k_y =0$) and large gravity velocities $C_0$. In case of the solar tachocline, however, these solutions are truncated by the Alfv\'{e}n wave branches for magnetic field strength $>10\, {\rm kG}$ \citep{Zaqarashvili2018a}.
\subsection{Damped wave dynamics}
We have now prepared the groundwork to discuss novel wave solutions under the effect of damping. Keeping the $\lambda$-dependent terms in Equation (\ref{eq:Wave}) but still neglecting the forcing potential $V$, we arrive at the wave equation 
\begin{align}
	\square^2_{v_A}v - C_0^2 \square_{v_A} \Delta v + (f^2 + \lambda^2) \frac{\partial^2 v}{\partial t^2} - C_0^2 \beta \frac{\partial}{\partial x}\frac{\partial v}{\partial t} \nonumber \\
	+2 \lambda \frac{\partial}{\partial t}\square_{v_A}v - \lambda C_0^2 \Delta \frac{\partial v}{\partial t}
	= 0. \label{eq:Damped}
\end{align} 
\subsubsection{Non-equatorial waves}
For non-equatorial waves, we can again perform a simple Fourier analysis $v \sim \exp(ik_x x + ik_y y - i\omega t)$, yielding the complex dispersion relation 
\begin{align}
	\omega^4 + 2i\lambda \omega^3- \left[2k_x^2 v_A^2 + f_0^2 + \lambda^2 + C_0^2(k_x^2 + k_y^2)\right]\omega^2 \nonumber \\
	-[C_0^2k_x \beta +2i\lambda k_x^2 v_A^2 + \lambda i C_0^2(k_x^2 + k_y^2)]\omega \nonumber \\
	+k_x^2 v_A^2 \left[k_x^2 v_A^2 + C_0^2(k_x^2 + k_y^2)\right] = 0. \label{eq:DispDampFull}
\end{align}
The attenuation of the waves is manifested in the imaginary part of the frequency $\omega$, which translates into an exponential decay of the Fourier modes. Although Equation (\ref{eq:DispDampFull}) appears quite delicate, the damping behavior turns out to be surprisingly simple for most wave modes. In the limit of high-frequency magneto-inertia waves, Equation (\ref{eq:DispDampFull}) approximates to 
\begin{align}
	\omega \approx \pm \sqrt{f_0^2 + C_0^2(k_x^2 + k_y^2) + 2k_x^2 v_A^2} -i\lambda. \label{eq:DampedInert}
\end{align}
The real part of Equation (\ref{eq:DampedInert}) is the same as in the inviscid case (\ref{Inert}), here Rayleigh frictions does not do anything except letting the waves decay exponentially, where the decay rate directly corresponds to the damping rate $\lambda$. For the slow frequency solutions in Equation (\ref{eq:DispDampFull}) in the limit of large-scale waves $k_x , k_y \rightarrow 0$ we find
\begin{align}
	\omega^2 + \frac{R_R^2\beta k_x + i\lambda R_R^2(k_x^2 + k_y^2) }{1 + \lambda^2/f_0^2 + R_R^2 (k_x^2 + k_y^2)}\omega \nonumber \\
	- \frac{R_R^2 v_A^2 k_x^2 (k_x^2 + k_y^2)}{1 + \lambda^2/f_0^2 + R_R^2 (k_x^2 + k_y^2)} = 0. \label{eq:DispSlowDamp}
\end{align}
In contrast to the inertia wave solution, damping is here also affecting the real part of $\omega$ by the $\lambda^2/f_0^2$ term in the denominator, effectively reducing the natural frequencies. However, this detuning effect is negligibly small if we focus on underdamped waves($\lambda < \omega$), where we readily find $\lambda / f_0 \ll 1$. If we further, as in section \ref{sec:FreeNonEq}, consider the radiative part of the tachocline, Equation (\ref{eq:DispSlowDamp}) simplifies to
\begin{align}
	\omega^2 + \frac{\beta k_x}{k_x^2 + k_y^2}\omega +i\lambda \omega - v_A^2 k_x^2 = 0 \label{eq:RossbyDamped}
\end{align}
as the analogue to Equation (\ref{eq:Rossby}). The hydrodynamic Rossby branch 
\begin{align}
	\omega \approx -\frac{k_x \beta}{k_x^2 + k_y^2} -i\lambda \label{eq:ClassicDamped}
\end{align}	
is in the same way affected by damping as magneto-inertia waves (\ref{eq:DampedInert}), the damping rate $\lambda$ is equal to the exponential decay rate. Intriguingly, we find a more complex and novel behavior for magneto-Rossby waves, underlying the dispersion relation 
\begin{align}
	\omega \approx \frac{k_x^2 v_A^2 (k_x^2 + k_y^2)}{k_x\beta + i\lambda (k_x^2 +k_y^2)} \nonumber \\
	=\frac{v_A^2 \beta k_x^3(k_x^2 + k_y^2)}{\beta^2 k_x^2 + \lambda^2(k_x^2 + k_y^2)^2} - \frac{i\lambda v_A^2 k_x^2 (k_x^2 + k_y^2)^2}{\beta^2 k_x^2 + \lambda^2(k_x^2 + k_y^2)^2}.
\end{align}
Since this solution is only valid for large scales $k_x R_0 \lesssim 1$, see \cite{Zaqarashvili2007}, we can again neglect the detuning term $\lambda^2 (k_x^2 + k_y^2)^2$, allowing to write the dispersion relation as follow
\begin{align}
	\omega = \omega_0 -i\lambda\frac{\omega_0^2}{\omega_A^2},
\end{align}
where $\omega_0$ is the inviscid natural frequency of slow magneto-Rossby waves (\ref{eq:MagnetoRossby}) and $\omega_A = \pm k_x v_A$ the Alfv\'{e}n frequency. This time, the decay rate differs from the damping rate by the factor $\omega_0^2 / \omega_A^2$. \cite{Zaqarashvili2007} have shown that the magneto-Rossby waves are always slower than Alfv\'{e}n waves (in contrast to hydrodynamic Rossby waves (\ref{eq:ClassicDamped})) such that we always have $\omega_0^2 / \omega_A^2 <1$ effectively reducing the decay rate. As a consequence, we can conclude that magneto-Rossby waves are more resilient to (linear) viscous damping than any other planetary wave modes, giving some more evidence that magneto-Rossby waves are indeed more relevant for the space weather dynamics than classic Rossby waves \citep{Dikpati2020a}. 

As a last remark, we want to stress the practical benefit of our approach to include damping. Treating Rayleigh's linear friction law as the only dissipation source is surely an oversimplification, but it can also be viewed as an empirical law since we know how it translates to decay rates in the different solutions. In principle, one can experimentally measure the exponential decay rate and translate it back to the damping rate $\lambda$ (e.g., by applying the factor $\omega_0^2 / \omega_A^2$ in the case of magneto-Rossby waves). This way, the model can correctly reflect all participating linear damping mechanisms and not only viscous damping. Such empirical approaches have been successfully implemented for simple free-surface wave systems \citep{Horstmann2020,Horstmann2021}. In these controllable mechanical systems it is, however, easy to determine decay rates, while it seems rather hopeless to extract them from space weather data, especially since solar Rossby waves are constantly and unpredictably excited. Nevertheless, one can at least roughly estimate the order of magnitude of $\lambda$, e.g., by measuring typical Rossby wave amplitudes (\cite{Mandal2020} determined amplitudes in the order of $1\, {\rm m}\, {\rm s}^{-1}$) and estimating the energy input, i.e., the energy stored in the differential rotation and toroidal magnetic field \citep{Dikpati2001}. Apart from that, it is worthwhile for future studies to investigate further dissipation mechanisms, be it Newtonian cooling \cite{Chang1982,Wu2001,Tsai2014} or magnetic damping, which have not yet been regarded in the framework of magnetohydrodynamic Rossby waves.   
\subsubsection{Equatorial waves}
At the equator, damped waves are more complex and different to non-equatorial waves since friction does not only alter the frequencies but also the meridional structure and phase of the wave profiles, as we will see in the following. In the vicinity of the equator, we have $f \approx \beta y$ and Equation (\ref{eq:Wave}) transforms into   
\begin{figure}[t!]
	\hspace*{-0.15cm}
	\includegraphics[scale=0.48]{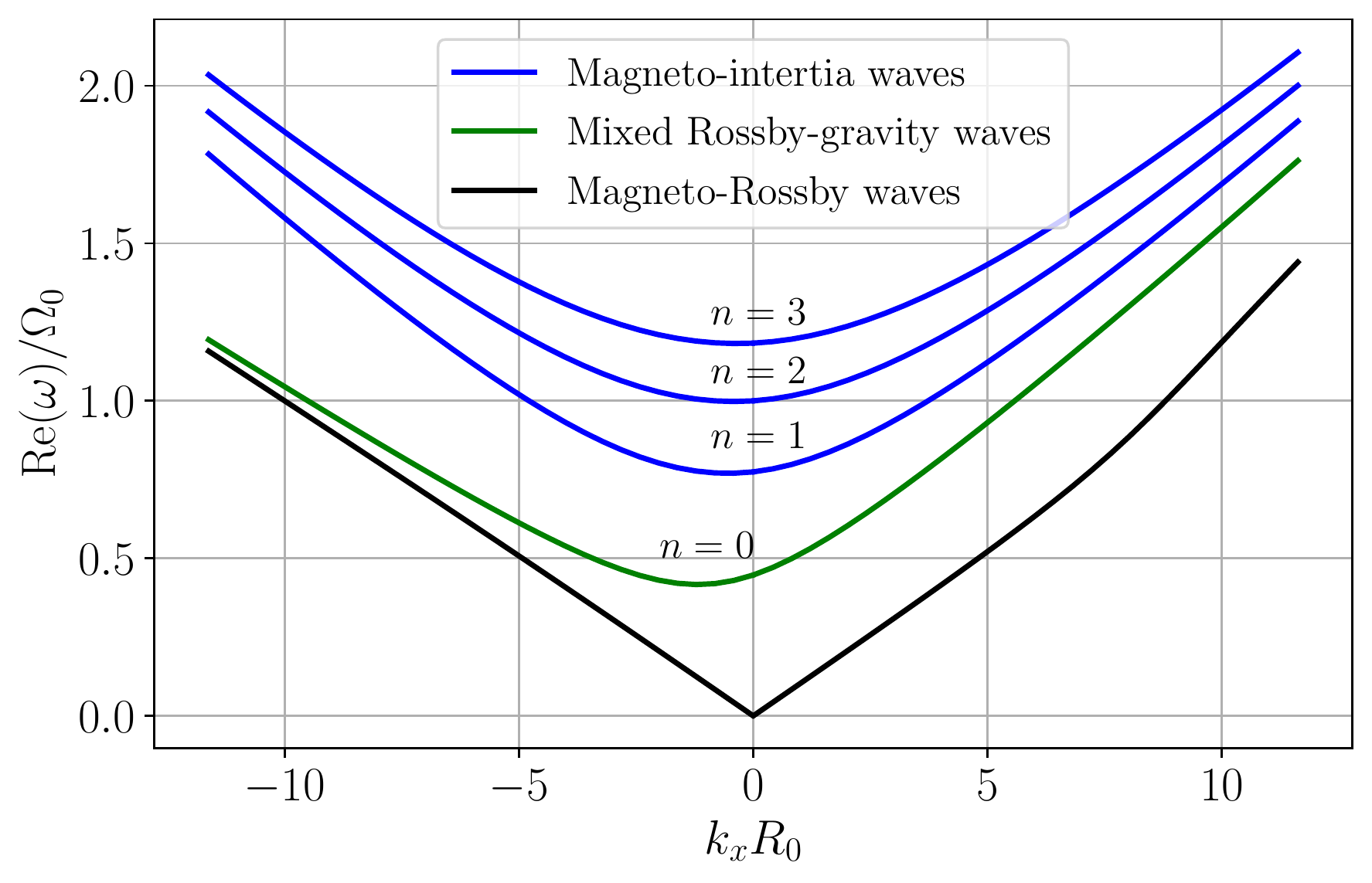}
	\caption{Dispersion curves of different wave modes calculated from equation (\ref{eq:DispEqDamp}) for the same parameters used by \cite{Zaqarashvili2018a}, i.e., $\Omega_0 = 26\cdot 10^{-7}\, {\rm s}^{-1}$, $R_0 = 5\cdot 10^{10}\, {\rm cm}$, $C_0 = 13 \cdot 10^{3}\, {\rm cm}\, {\rm s}^{-1}$, $v_A = 12.6 \cdot 10^{3}\, {\rm cm}\, {\rm s}^{-1}$, $\beta = 1.04 \cdot 10^{-16}\, {\rm cm}^{-1}\, {\rm s}^{-1}$ under the action of the damping rate $\lambda = 0.01\Omega_0$. The blue curves correspond to magneto-inertia waves, the green curve to a mixed magneto-Rossby-gravity wave solution and the black curve describes the magneto and hydrodynamic Rossby waves with $n=1$.}
	\label{fig:FrequenciesEq}
\end{figure}
\begin{align}
	\square^2_{v_A}v - C_0^2 \square_{v_A} \Delta v + (\beta^2 y^2 + \lambda^2) \frac{\partial^2 v}{\partial t^2} - C_0^2 \beta \frac{\partial}{\partial x}\frac{\partial v}{\partial t} \nonumber \\
	+2 \lambda \frac{\partial}{\partial t}\square_{v_A}v - \lambda C_0^2 \Delta \frac{\partial v}{\partial t}
	= 0 \label{eq:DampedEq}
\end{align} 
for $V=0$. By using the same ansatz as before, $v = v_y(y)\exp(ik_x x - i\omega t)$, one readily gets the determining equation
\begin{align}
	\frac{{\rm d}^2 v_y}{{\rm d} y^2} + \left[\frac{\omega^2 + i\lambda \omega - k_x^2 v_A^2}{C_0^2} - k_x^2 \right. - \nonumber \\ \left. -\frac{k_x \beta \omega}{\omega^2 +i\lambda \omega - k_x^2 v_A^2} - \mu^2 y^2 \right] v_y = 0, \label{eq:CylindricalDamped}
\end{align}
where
\begin{align}
	\mu = \frac{\beta \omega}{C_0\sqrt{\omega^2 + i\lambda \omega - k_x^2 v_A^2}}. \label{eq:Mu}	
\end{align}
The dispersion relation follows again from the solvability condition (\ref{eq:Contraint}):
\begin{align}
	(\omega^2 + i\lambda \omega - k_x^2 v_A^2)(\omega^2 + i\lambda \omega - k_x^2(C_0^2 + v_A^2)) - k_x \beta C_0^2 \omega \nonumber \\
	=  \beta C_0|\omega|\sqrt{\omega^2 + i\lambda \omega - k_x^2 v_A^2}(2n+1). \label{eq:DispEqDamp}
\end{align}  
The complex-valued square root on the right-hand side highly complicates further analysis. It is no longer feasible to derive analytical approximations for the different wave types. Therefore, we analyze the dispersion relation numerically, which is a bit delicate since the square root introduced spurious solutions. In order to get around of this issue we squared both sides of the dispersion relation\textemdash for the price of introducing four invalid solutions\textemdash and determined the eight zeros of the resulting polynomial by calculating the eigenvalues of the companion matrix. Afterwards, we checked the validity of the solutions to guarantee that they fulfill the initial dispersion relation (\ref{eq:DispEqDamp}) . Figure \ref{fig:FrequenciesEq} shows the wave frequencies Re$(\omega)$ as a function of the wave numbers $k_x$ on the example of the solar tachocline with the same parameters as used by \cite{Zaqarashvili2018a} under the effect of a hypothetical friction parameter of $\lambda = 0.01\Omega_0$.  
\begin{figure}[t!]
	\hspace*{-0.3cm}
	\includegraphics[scale=0.48]{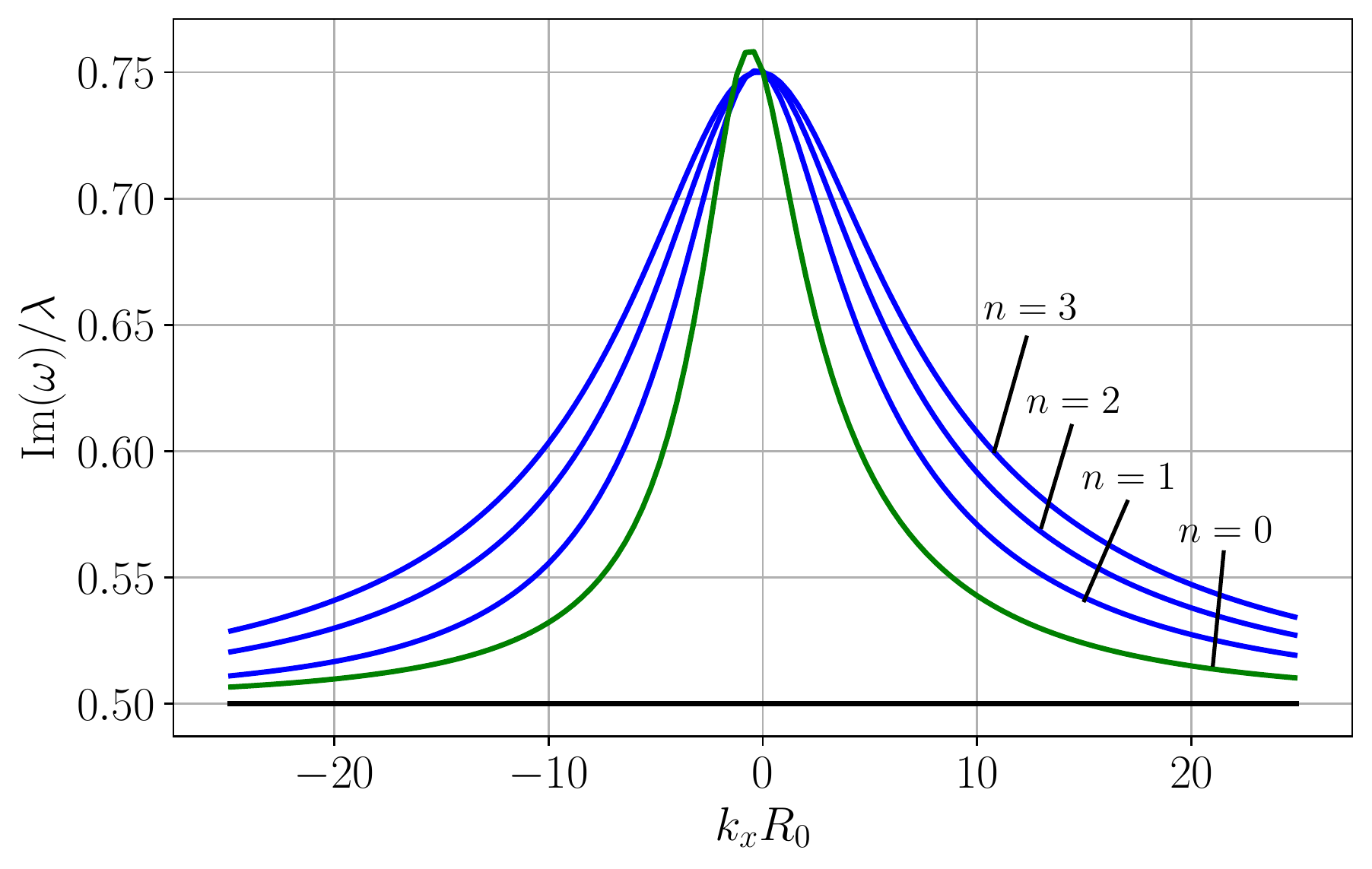}
	\caption{Decay rates ${\rm Im}(\omega)$ corresponding to the frequencies shown in figure \ref{fig:FrequenciesEq}. The blue curves correspond to magneto-inertia waves, the green curve to a mixed magneto-Rossby-gravity wave solution and the black curve describes the magneto and hydrodynamic Rossby waves with $n=1$.}
	\label{fig:DampingRates}
\end{figure}
When compared to the undamped solutions, no difference can be discerned for all inertia, gravity and Rossby wave solutions. We found that the effect of $\lambda$ on the eigenfrequencies is entirely negligible except for large friction coefficients in the order of $\lambda \sim |\omega|$. The corresponding decay rates, however, show a more complex behavior than the non-equatorial counterparts. Figure \ref{fig:DampingRates} shows  the imaginary part of the frequency Im$(\omega)$ of the same solutions. It can be recognized that the resulting damping rates are smaller than for non-equatorial waves under the same conditions. The decay behavior  of magneto-inertia waves shows an interesting length-scale dependence. Large-scale Poincar\'{e} waves ($k_x \rightarrow 0$) diminish with Im$(\omega) \approx 0.75 \lambda$, whereas the small-scale representatives approach the smaller decay rate of Im$(\omega) = 0.5\lambda$ for $k_x \rightarrow \pm \infty$. Furthermore, it can be observed that higher wave modes tend to dissipate more rapidly. Rossby waves always decay with Im$(\omega) = 0.5\lambda$ independently of the wavelength, which is half as fast as non-equatorial Rossby waves, see Equation (\ref{eq:ClassicDamped}). This can be explained by the fact that Rossby waves are bordered by Alfv\'{e}n waves and have very similar eigenfrequencies \citep{Zaqarashvili2018a}. The frequency of Alfv\'{e}n waves, however, is governed by the square root term of equation (\ref{eq:DispEqDamp}) $\omega^2 + i\lambda \omega - k_x^2 v_A^2 = 0$, yielding a decay rate of Im$(\omega) = -\lambda/2$ in agreement to the magneto-Rossby waves.
\begin{figure}[t!]
	\hspace*{-0.3cm}
	\includegraphics[scale=0.48]{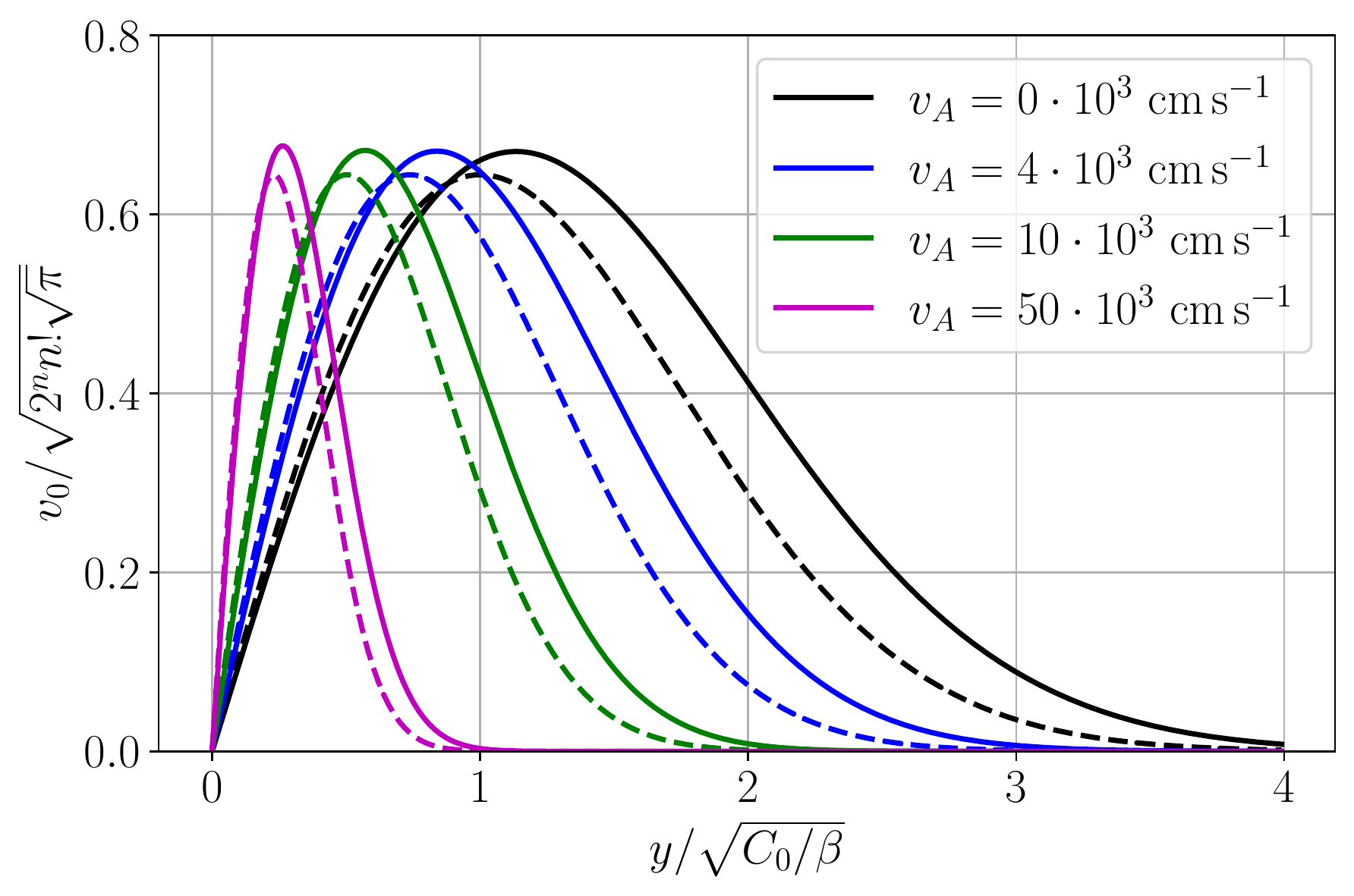}
	\caption{Meridional distributions of $n=1$ Rossby waves in the equatorial waveguide at $\omega = -1\cdot 10^{-7}\, {\rm s} \equiv k_x R = 2$ for different Alfv\'{e}n velocities $v_A$. The latitude coordinate $y$ is normalized by the inviscid and non-magnetic Rossby scale $\sqrt{C_0 / \beta}$. The dashed lines correspond to inviscid waves and the solid lines to damped waves with a friction coefficient of $\lambda = |\omega|$.}
	\label{fig:Scale1}
\end{figure}

\begin{figure}[t!]
	\hspace*{-0.3cm}
	\includegraphics[scale=0.48]{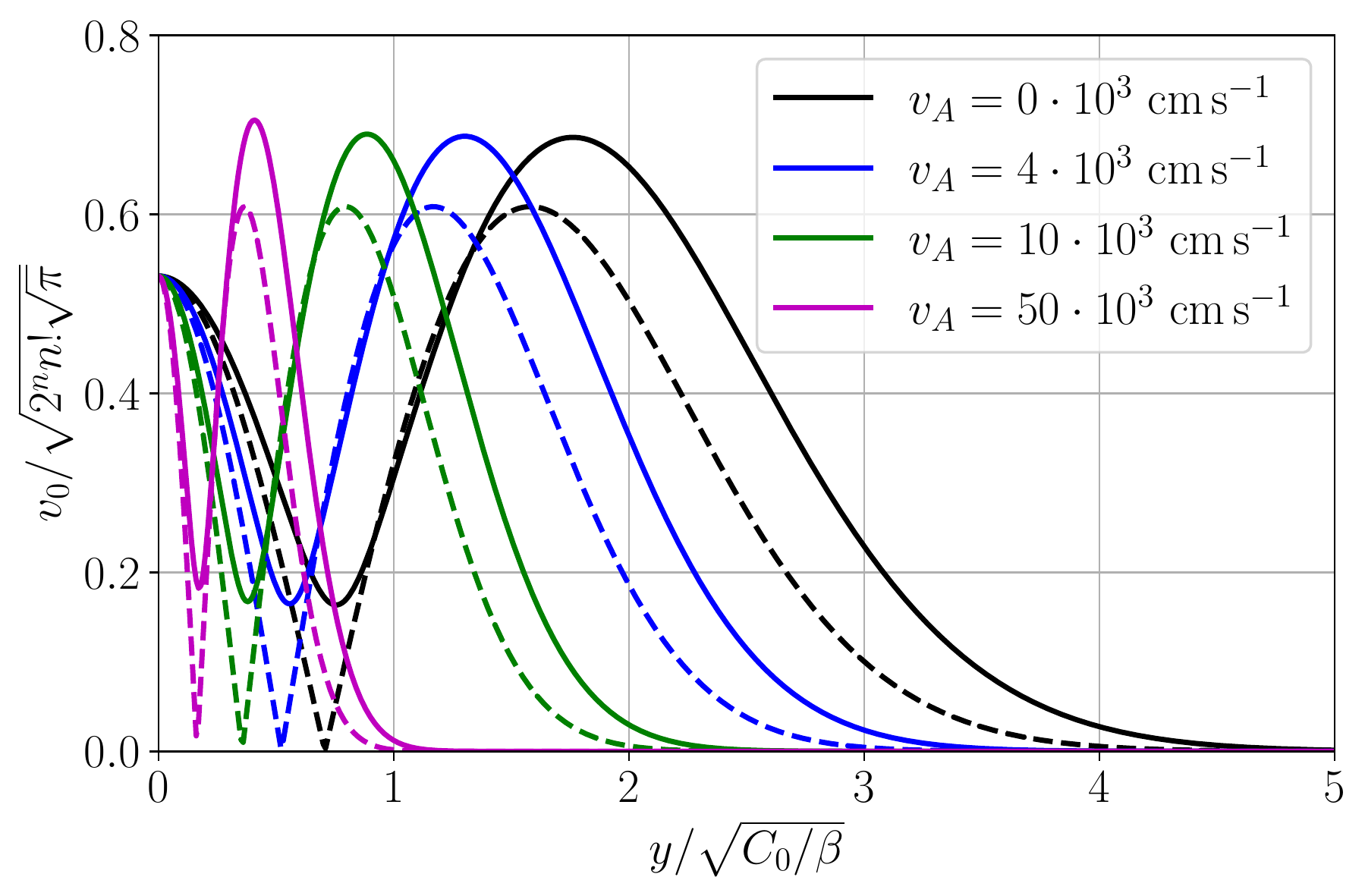}
	\caption{Meridional distributions of $n=2$ Rossby waves in the equatorial waveguide at $\omega = -1\cdot 10^{-7}\, {\rm s} \equiv k_x R = 2$ for different Alfv\'{e}n velocities $v_A$. The latitude coordinate $y$ is normalized by the inviscid and non-magnetic Rossby scale $\sqrt{C_0 / \beta}$. The dashed lines correspond to inviscid waves and the solid lines to damped waves with a friction coefficient of $\lambda = |\omega|$.}
	\label{fig:Scale2}
\end{figure}

So far we have analyzed damped waves which decrease in time. In many cases, as for some forced wave systems, it is also interesting to study free wave motions under the effect of damping. Free waves always have real frequencies, such that the dispersion relation (\ref{eq:DispEqDamp}) then necessarily introduces complex wave numbers $k_x$ for $\lambda > 0$, letting the waves decay spatially in the zonal direction. Such waves can evolve under spatially limited but constant forcing, whereby the wave amplitudes are maintained in the region of forcing and spatially fade away outside. In geophysics it is well known that friction can modify the meridional scales and introduce a phase shift in the meridional structure of planetary waves \citep{Mofjeld1981,Yamagata1985}. Therefore, the question arises to what extent meridional scales of magneto-Rossby waves can be affected. For real frequencies, the wave number of slow and large-scale magneto-Rossby waves can be approximated as follows:
\begin{align}
	k_x^2 = \frac{(2n+1)^2 (\omega^2 +i\lambda \omega)}{C_0^2 + v_A^2 (2n+1)^2}. \label{eq:Kx}
\end{align}
We use this dispersion relation for the following calculations. 
Figures \ref{fig:Scale1} and \ref{fig:Scale2} show normalized meridional velocity profiles $v(y)$ of $n=1$ and $n=2$ magneto-Rossby waves for different  Alfv\'{e}n velocities $v_A$ without (dashed lines) and with strong damping $\lambda = \omega$ (solid lines) in accordance to \cite{Mofjeld1981}. Two opposing tendencies can be observed. As with classic Rossby waves, friction always broadens the wave profiles. However, this effect is increasingly counterbalanced by the presence of toroidal magnetic fields that tight the magneto-Rossby wave profiles. This was to be expected since the oscillatory scale (\ref{eq:LatExt}) more and more decreases as magneto-Rossby frequencies approach Alfv\'{e}n wave frequencies for $v_A \rightarrow \infty$. Altogether, it can be concluded that the meridional scale variation due to friction is negligible relative to the magnetic contribution. Finally, we can also see that friction causes larger absolute latitudinal velocities (but smaller zonal velocities, not shown here). This is in conformity with the observations by \cite{Mofjeld1981}, there is no significant difference between classic and magnetic Rossby waves.

\subsection{Forced wave dynamics}
Building on these preliminary investigations, we can now turn on to the study of forced wave solutions.
For the sake of keeping the analysis as simple, we focus on wave responses to a single tide generating planet of mass $M_t$ moving at a fixed distance $r$ with the angular frequency $\Omega_t$ around a rotating star within the equatorial plane (zero declination). We show in appendix \ref{sec:Tidal} that the resulting leading-order tidal potential acting in the $\beta$-plane approximates to
\begin{align}
	V= K\left(\frac{1}{2} +\frac{y}{R_0}\right)\left[1+ \cos\left(\frac{2x}{R_0} - 2(\Omega_t -\Omega_0)t\right)\right] \label{eq:PotNon}
\end{align}
at mid-latitudes $\phi_0 = 45^{\circ}$ and to
\begin{align}
	V= K\left[1 + \cos\left(\frac{2x}{R_0} - 2(\Omega_t -\Omega_0)t\right)\right] \label{eq:PotEq}
\end{align} 
in the vicinity of the equator $\phi_0 = 0^{\circ}$. The forcing amplitude is given as
\begin{align}
	K = \frac{3}{4}\frac{G M_t}{R_0}\left(\frac{R_0}{r}\right)^3
\end{align}
with $G$ referring to the gravitational constant. Two characteristic forcing frequencies appear in the potentials; the rotation frequency of the star $\Omega_0$ and an external frequency $\Omega_t$ dictated by the tide generating planet. If we consider the Sun exposed to the tidally dominant planet Jupiter, we find very different periods of $T_0 \approx$ 25 days and of $T_t \approx 11$ years, so that, in principle, a high bandwidth of planetary waves can be excited ranging from high frequency magneto-inertia to very slow magneto-Rossby waves. In the following sections, we present explicit solution of wave responses under resonant and non-resonant forcing conditions for both unbounded and equatorially trapped waves. For didactic considerations we start this time with equatorial waves.

\subsubsection{Equatorial waves} \label{sec:ForcedEq}
Inserting the equatorial potential (\ref{eq:PotEq}) into the wave equation (\ref{eq:Wave}) yields 
\begin{align}
	\square^2_{v_A}v - C_0^2 \square_{v_A} \Delta v + \beta^2 y^2 \frac{\partial^2 v}{\partial t^2} - C_0^2 \beta \frac{\partial}{\partial x}\frac{\partial v}{\partial t} \nonumber \\
	+2 \lambda \frac{\partial}{\partial t}\square_{v_A}v - \lambda C_0^2 \Delta \frac{\partial v}{\partial t} + \lambda^2 \frac{\partial^2 v}{\partial t^2}  = \beta y\frac{\partial}{\partial x}\frac{\partial^2 V}{\partial t^2} \nonumber \\
	= \beta y\frac{8K\Omega^2}{R_0} \sin\left(\frac{2x}{R_0} - 2\Omega t\right).
	\label{eq:WaveEq}
\end{align} 
\begin{figure*}[t!]
	\includegraphics[scale=0.5]{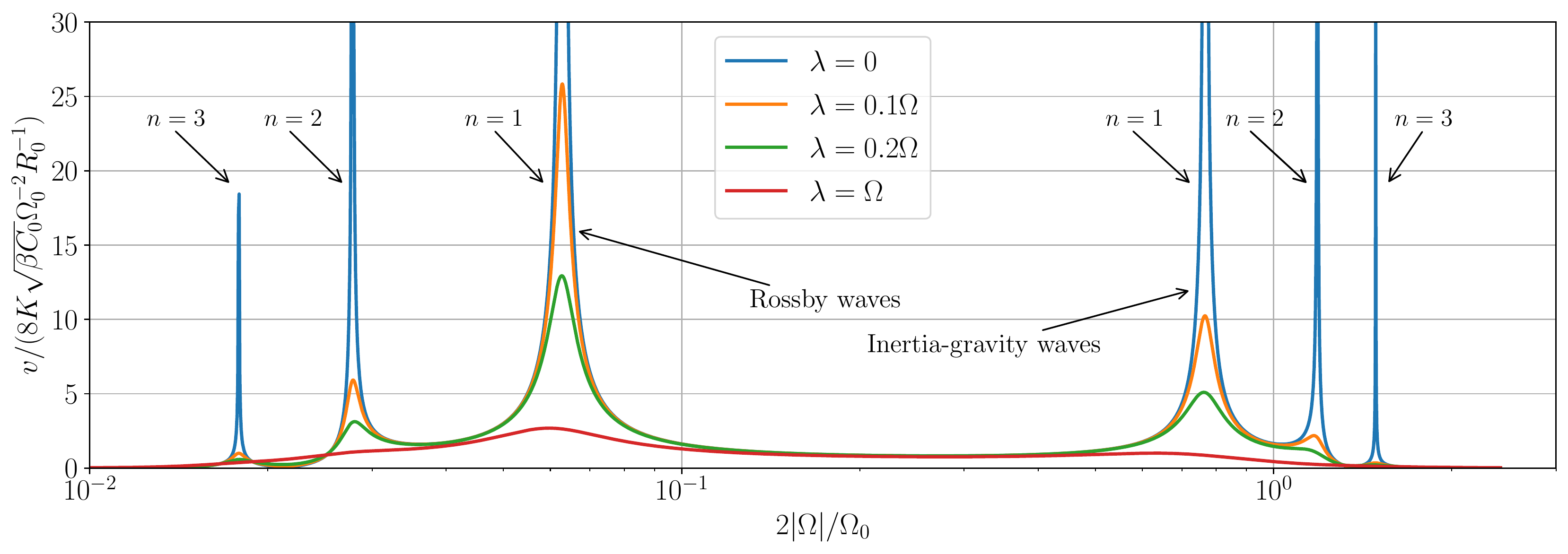}
	\includegraphics[scale=0.5]{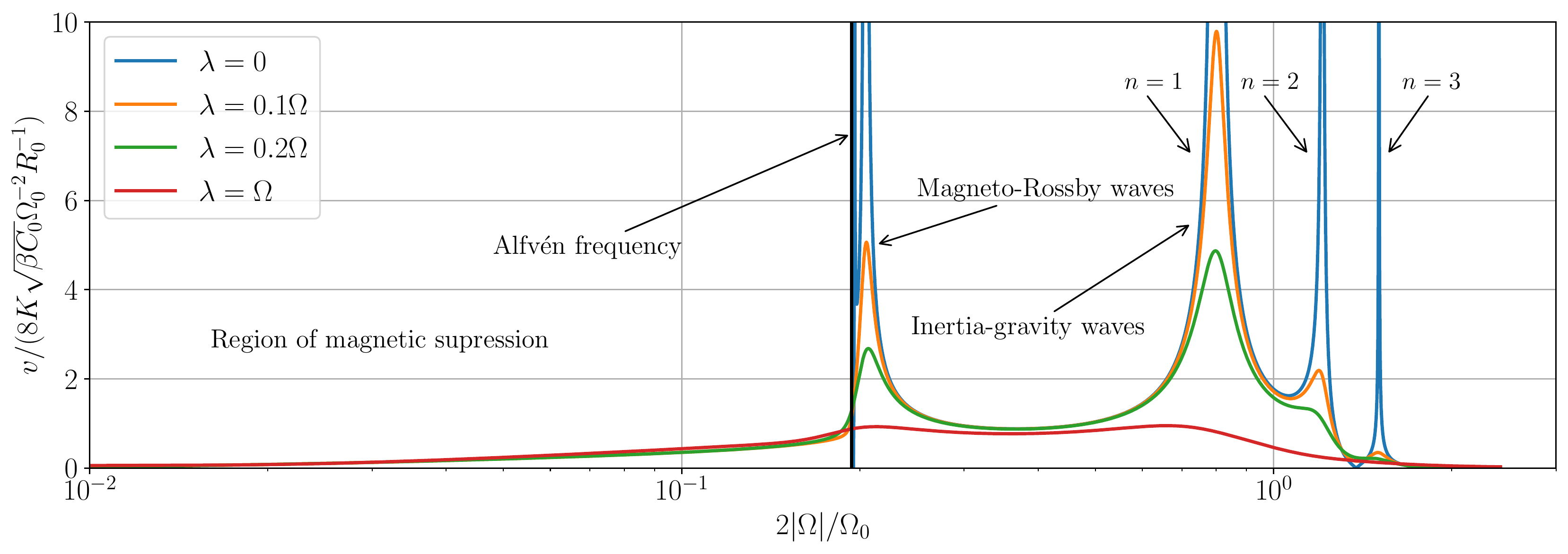}
	\caption{Normalized velocity amplitudes due to solution (\ref{eq:SolEq}) for the hydrodynamic $v_A = 0\, {\rm cm}\, {\rm s}^{-1}$ (top) and magnetic case $v_A = 12.6\cdot 10^3 \, {\rm cm}\, {\rm s}^{-1}$ (bottom) as a function of the normalized forcing frequency $\Omega$. The calculated frequency range includes the first three retrograde propagating magneto-Rossby and magneto-inertia wave modes. We used  $\Omega_0 = 26\cdot 10^{-7}\, {\rm s}^{-1}$, $R_0 = 5\cdot 10^{10}\, {\rm cm}$, $C_0 = 13 \cdot 10^{3}\, {\rm cm}\, {\rm s}^{-1}$, $\beta = 1.04 \cdot 10^{-16}\, {\rm cm}^{-1}\, {\rm s}^{-1}$.}
	\label{fig:Forced}
\end{figure*}	
The forcing frequencies were unified to $\Omega_t -\Omega_0 = \Omega$ for the sake of clarity. Since the equatorial potential is independent of the latitude $y$ only the Coriolis forcing term $\beta y \partial_x \partial^2_t V$ remains such that the leading order tidal action is affected neither by the Alfv\'{e}n speed $v_A$ nor by damping $\lambda$, and therefore manifests itself in the same manner as with classic Rossby waves. The Coriolis parameter $f \approx \beta y$ reintroduces the $y$ coordinate, which requires us to expand $y$ as an orthogonal series of parabolic functions:
\begin{align}
	y=\sum_{n=1}^{\infty}\frac{2^{\frac{5}{2}-2n}}{\sqrt{\mu}(n-1)!}\exp\left(-\frac{\mu y^2}{2}\right)H_{2n-1}(\sqrt{\mu}y). \label{eq:Series}
\end{align}
We have used the practical parabolic orthogonality condition 
\begin{align}
	\int_{-\infty}^{\infty}\exp(-\mu y^2)H_n(\sqrt{\mu}y)H_m(\sqrt{\mu}y){\rm d}y = 2^n n! \sqrt{\frac{\pi}{\mu}}\delta_{nm}
\end{align}
for the expansion. As a remarkable intermediate result, Equation (\ref{eq:Series}) implies that only uneven wave modes $n \rightarrow 2n-1$ can respond to tidal forcing in the leading order, which are always antisymmetric around the equator.
We show in appendix \ref{sec:SolEq} that the forced wave problem can be solved explicitly by introducing the ansatz 
\begin{equation}
	v = \sum_{m=0}^{\infty}\sum_{n=0}^{\infty}\alpha_{m,n}(t) \exp\left(i\frac{mx}{R_0}\right)\exp\left(-\frac{\mu y^2}{2}\right)H_n (\sqrt{\mu}y), \label{eq:Ansatz}
\end{equation} 
where $\alpha_{m,n}(t)$ are modal coefficients to be determined. The resulting solution can be expressed as an infinite series of parabolic functions:
\begin{widetext}
\vspace*{-0.45cm}
\begin{align}
&v(x,y,t) = \sum_{n=1}^{\infty}\Lambda_n \exp\left(i\frac{2x}{R_0} - i2\Omega t\right)\exp\left(-\frac{\mu y^2}{2}\right)H_{2n-1}(\sqrt{\mu}y), \label{eq:SolEq} \\
&{\rm with} \ \ \ \mu = \frac{\beta |2\Omega|}{C_0\sqrt{4\Omega^2 + 2i\lambda \Omega -  \frac{4v_A^2}{R_0^2}}} \ \ \ {\rm and} \nonumber \\
&\Lambda_n = \frac{8iK\beta \Omega^2 2^{\frac{5}{2}-2n}(R\sqrt{\mu}(n-1)!)^{-1}}{\left(4\Omega^2 +2i\lambda \Omega - \frac{4 v_A^2}{R_0^2}\right)\left(4\Omega^2 + 2i\lambda \Omega -\frac{4}{R_0^2}(C_0^2 + v_A^2)\right) - C_0^2\frac{4\beta}{R_0}\Omega - C_0 \beta|2\Omega|\sqrt{4\Omega^2 +2i\lambda \Omega - \frac{4v_A^2}{R_0^2}}(4n-1)}. \nonumber
\end{align}
\end{widetext}
Solution (\ref{eq:SolEq}) consists of two parts. A real part $\sim \text{Re}[\Lambda_n]$ describing non-resonant forced waves between the different wave modes and an imaginary part $\sim \text{Im}[\Lambda_n]$ capturing resonant waves in the vicinity of the eigenfrequencies described by the dispersion relation (\ref{eq:DispDampFull}) for $\omega = 2\Omega$. The velocity amplitude scales with $\sim K \sqrt{\beta C_0}$, showing that pure Alfv\'{e}n waves ($\beta = C_0 = 0$) do not respond to tidal forcing at the equator. Besides, we would like to draw attention to the fact that (\ref{eq:SolEq}) is formally only a specific solution of Equation (\ref{eq:WaveEq}), the general solution is composed of the specific and homogeneous solution. The homogeneous part, however, only describes transient phases, e.g., in classical sloshing experiments the initial wave motions taking place directly after switching on the shaking table, and decays exponentially after some settling time. After the transient phase the saturated, quasi-steady wave responses are fully governed by the specific solution presented here.   	

For deeper insights into the dynamics of forced planetary waves, we have calculated the wave responses as a function of the forcing frequency $\Omega$ within a range comprising the first three (magneto)-Rossby and (magneto)-inertia wave modes $n=1,2,3 \rightarrow H_1 , H_3 , H_5$. The upper plot in Figure \ref{fig:Forced} shows normalized wave amplitudes of classical planetary waves $v_A = 0$ for different dynamic damping coefficients $\lambda$ related to the forcing frequency $\Omega$, so that $\lambda$ always references to the characteristic time in which equivalent unforced waves would decay. For example, the $\lambda = 0.2\Omega$ corresponds to a damping rate of $0.1\Omega$, see the Rossby wave solution in Figure \ref{fig:DampingRates}, which would let the wave diminish substantially within around ten periods following an artificial elimination of the excitation force. Figure \ref{fig:Forced} reveals that the first Rossby mode $n=1$ is excited most strongly, about five times stronger than the first inertia-gravity wave and up to 25 times stronger (for $\lambda = 0.1\Omega$) than non-resonant waves occurring in the frequency band between the low-frequency Rossby and high-frequency inertia wave solutions. Further it can be recognized that the peak amplitudes decrease rapidly along with increasing wave numbers showing that the large-scale modes are always most significant for tidal interactions. 
As another important general result, it is evident from the resonance curves that different planetary waves will respond to different classes of planet-hosting stars. Is the star's rotation far higher than the planet's orbit frequency $|\Omega_0| \gg |\Omega_t|$, as it is the case for our Sun forced by Jupiter, only fast inertia-gravity waves will be exited. Are both angular frequency very close to each other (but not equal), as often the case for stars hosting tidally-locked planets or also for binary stars, slow Rossby modes are expected to be stimulated instead. This result is only true for stars hosting one single (significant) tide-generating planet. If there are several planets involved, Rossby waves can still be excited by low-frequency alignment periodicities, e.g., spring tides, visible in the envelope of the combined tidal potentials, even if we have  $|\Omega_0| \gg |\Omega_t|$ for all participating planets, see section \ref{sec:Plau}.
	
For comparison, the lower plot in Figure \ref{fig:Forced} shows the same frequency spectrum but for magnetic planetary waves with $v_A = 12.6\cdot 10^3\, {\rm cm}\, {\rm s}^{-1}$. In accordance with \cite{Zaqarashvili2018a}, it can be seen that Rossby-waves below the Alfv\'{e}n frequency (vertical black line) are largely suppressed. Magneto-Rossby waves accumulate closely above the Alfv\'{e}n frequency, where they respond with much smaller amplitudes. The inertia waves, in contrast, remain largely unaffected. As a novel and rather striking result, we can also see that damping allows the waves to overcome the Alfv\'{e}n frequency barrier. Inviscid planetary waves below $\omega = \pm 2 v_A /R_0$ do not exists, whereas damped waves are still able to respond at frequencies far below $2|\Omega|/\Omega_0 = 10^{-1}$. This behavior is intriguing, but it must be noted that constant toroidal magnetic fields generally inhibit Rossby waves and antagonize the tidal response, making these waves unlikely to be of any practical importance. At this point, it might be of interest for future studies to analyze the tidal response of Rossby waves subject to more realistic nonuniform and also oscillatory magnetic fields, which are not exposed to a cut-off frequency \citep{Zaqarashvili2018a}.
\subsubsection{Non-equatorial waves} \label{sec:MidResponse}
We proceed with solving the forced wave problem at mid-latitudes, which is a bit more intricate insofar that here the tidal potential (\ref{eq:PotNon}) depends, in contrast to the equatorial potential (\ref{eq:PotEq}), on the local latitude $y$ so that all three potential terms in Equation (\ref{eq:Wave}) must be taken into account. We obtain the following forced wave equation:
\begin{widetext} 
\begin{align}
	\square^2_{v_A}v - C_0^2 \square_{v_A} \Delta v + f_0^2 \frac{\partial^2 v}{\partial t^2} - C_0^2 \beta \frac{\partial}{\partial x}\frac{\partial v}{\partial t} 
	+2 \lambda \frac{\partial}{\partial t}\square_{v_A}v - \lambda C_0^2 \Delta \frac{\partial v}{\partial t} + \lambda^2 \frac{\partial^2 v}{\partial t^2} 
	 = f_0\frac{\partial}{\partial x}\frac{\partial^2 V}{\partial t^2} - \lambda\frac{\partial}{\partial y}\frac{\partial^2 V}{\partial t^2}  - \frac{\partial}{\partial t}\frac{\partial}{\partial y}\square_{v_A}V \nonumber \\
	= \left[f_0 \Omega +2 \Omega^2 - \frac{2v_A^2}{R_0^2} + \frac{2f_0 \Omega}{R_0}y\right] \frac{4K\Omega}{R_0}\sin\left(\frac{2x}{R_0} - 2\Omega t\right) + \frac{4K\lambda \Omega^2}{R_0}\cos\left(\frac{2x}{R_0} - 2\Omega t\right)
	\label{eq:WaveNonEq}
\end{align} 
\end{widetext}
Both constant and $y$-proportional terms remain on the right-hand side of (\ref{eq:WaveNonEq}), which moreover involves different phases $\sim \sin(2x/R_0 - 2\Omega t)$ and $\sim \cos(2x/R_0 - 2\Omega t)$. Hence, we are required to expand two different Fourier series for constant terms and $y$, which, however, gives rise to the difficulty that mid-latitude waves are meridionally unbounded so that the wave numbers $k_y$ are in principle arbitrary. Hence, we need to constrain the meridional scale in a purposeful way. On the one hand, we are interested in large-scale responses, but the $\beta$-plane approximation becomes more and more inaccurate with increasing meridional dimensions on the other hand. As the best compromise, we confine the wave problem into a waveguide defined by the interval $-R_0 /2 \leq y \leq R_0 /2$, with $y= 0$ defining the latitude $\phi_0 = 45^{\circ}$. In spherical coordinates, this interval corresponds approximately to the meridional band $15^{\circ} \lesssim \phi_0 \lesssim 75^{\circ}$, the respective latitudinal range of $60^{\circ}$ is widely regarded as the non-equatorial $\beta$-plane limit. Within the interval $-R_0 /2 \leq y \leq R_0 /2$, we can describe the forcing terms as Fourier series by expanding 
\begin{equation}
	1 = \sum_{n=1}^{\infty}\frac{4}{(2n-1)\pi}(-1)^{n-1}\cos\left(\frac{(2n-1)\pi y}{R_0}\right) \label{eq:FSeries}
\end{equation}
and
\begin{align}
	y = \sum_{n=1}^{\infty}\frac{4R_0}{(2n-1)^2\pi^2}(-1)^{n-1}\sin\left(\frac{(2n-1)\pi y}{R_0}\right). \label{eq:FSeries2}
\end{align}
Interestingly, both series only involve uneven latitudinal wave modes $2n-1$, which are the only modes that can respond in the leading order. Hence, the tidal potential, although manifested very differently nearby to and far away from the equator, imposes the same symmetry on mid-latitude waves as it does on equatorial waves. The forced wave problem can now be solved most easily by inserting the ansatz 
\vspace*{-0.2cm}
\begin{align}
	v &= \sum_{m=0}^{\infty}\sum_{n=0}^{\infty}\left[\alpha_{m,n}(t)\sin\left(\frac{m}{R_0}x\right) + \beta_{m,n}(t)\cos\left(\frac{m}{R_0}x\right) \right] \nonumber \\
	&\times \exp\left(\frac{in\pi}{R_0}y\right) \label{eq:Ansatzb}	
\end{align}
into Equation (\ref{eq:WaveNonEq}) as we show in Appendix \ref{sec:SolNonEq}. Since the right-hand side forcing terms in Equation (\ref{eq:WaveNonEq}) contains terms both symmetric and antisymmetric in the ($x,t$) space, we need two independent modal coefficients $\alpha_{m,n}(t)$ and $\beta_{m,n}(t)$ to solve the problem. We find the following explicit analytic solution: 
\begin{widetext}
	\begin{align}
		v(x,y,t) &= \sum_{n=1}^{\infty} \frac{\bar{K}_2+i\bar{K}_1}{2\left(16\Omega^4-4\Omega^2\bar{\Lambda}_1+\bar{\Lambda}_3-\frac{4\beta C_0^2\Omega}{R_0}\right)-4i\Omega(8\lambda\Omega^2-\bar{\Lambda}_2)}\exp\left(2i\Omega t-2i\frac{x}{R_0}+\frac{i(2n-1)\pi}{R_0}y \right) \nonumber\\
		&+\sum_{n=1}^{\infty} \frac{\bar{K}_2-i\bar{K}_1}{2\left(16\Omega^4-4\Omega^2\bar{\Lambda}_1+\bar{\Lambda}_3-\frac{4\beta C_0^2\Omega}{R_0}\right)+4i\Omega(8\lambda\Omega^2-\bar{\Lambda}_2)}\exp\left(-2i\Omega t+2i\frac{x}{R_0}+\frac{i(2n-1)\pi}{R_0}y \right). \label{eq:SolMid} 
\end{align}
\begin{figure*}
	\includegraphics[scale=0.5]{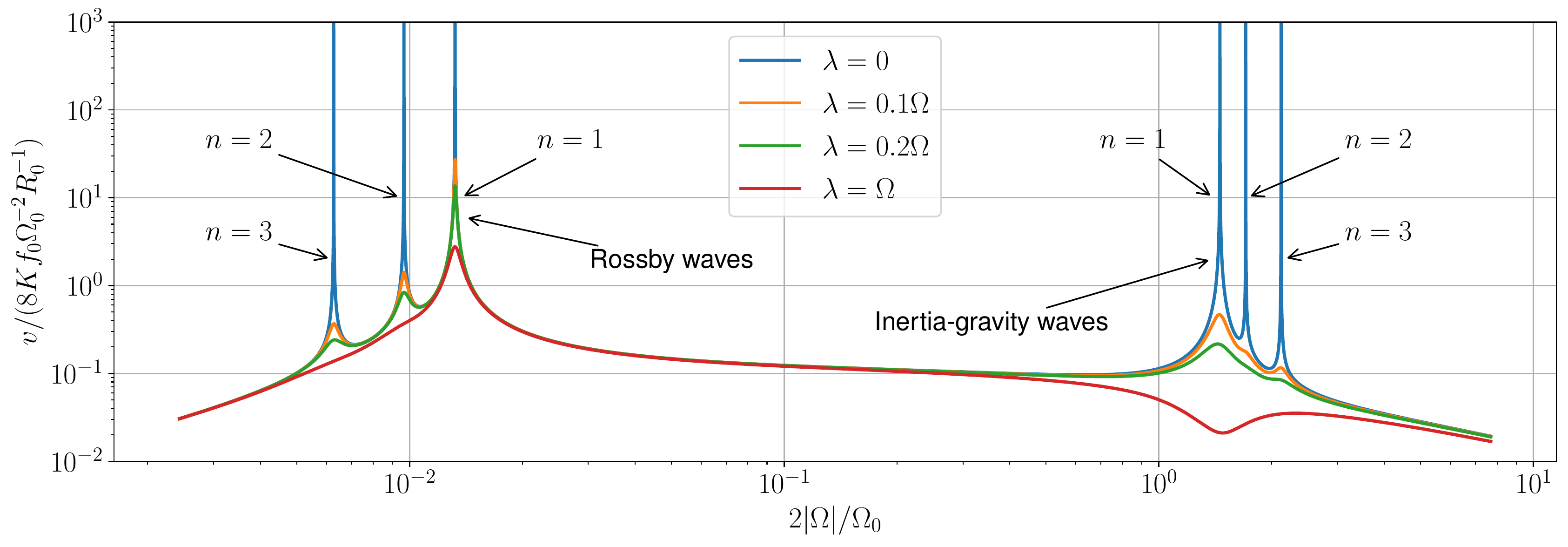}
	\includegraphics[scale=0.5]{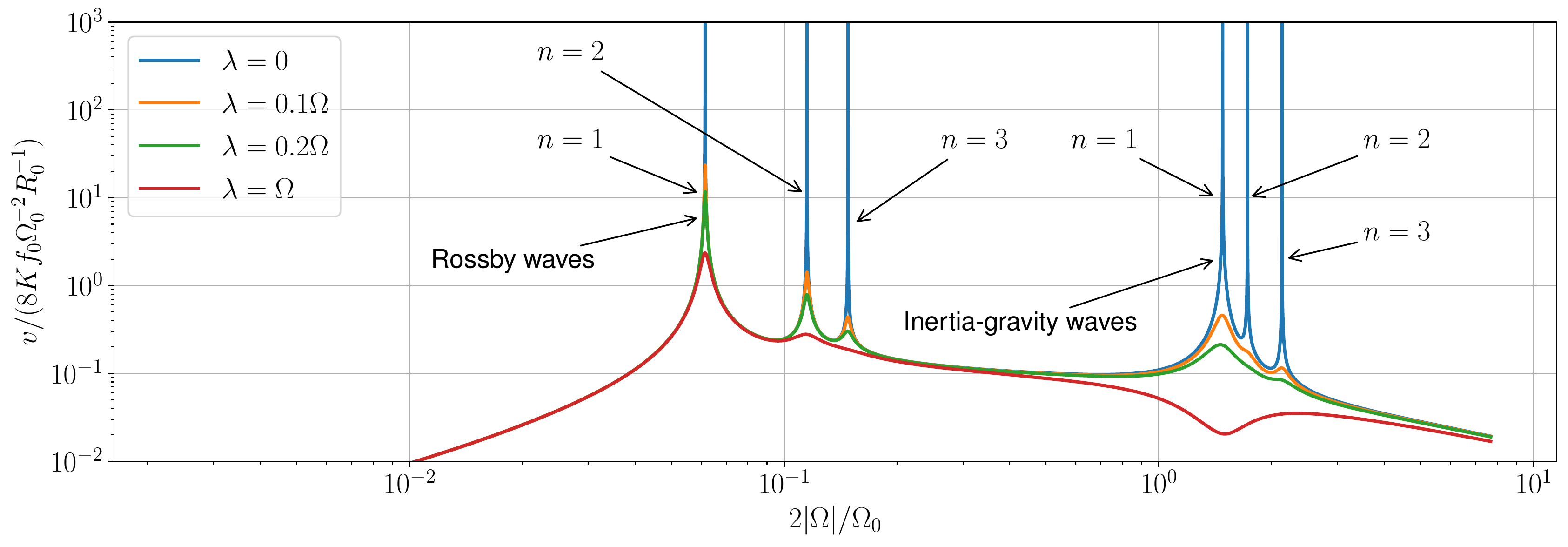}
	\caption{Normalized velocity amplitudes due to solution (\ref{eq:SolMid}) for $v_A = 0\, {\rm cm}\, {\rm s}^{-1}$ (top) and $v_A = 12.6\cdot 10^3 \, {\rm cm}\, {\rm s}^{-1}$ (bottom) as a function of the normalized forcing frequency $\Omega$. The calculated frequency range includes the first three retrograde propagating magneto-Rossby and magneto-inertia wave modes. We used  $\Omega_0 = 26\cdot 10^{-7}\, {\rm s}^{-1}$, $R_0 = 5\cdot 10^{10}\, {\rm cm}$, $C_0 = 13 \cdot 10^{3}\, {\rm cm}\, {\rm s}^{-1}$, $\beta = 1.04 \cdot 10^{-16}\, {\rm cm}^{-1}\, {\rm s}^{-1}$.}
	\label{fig:ForcedRetro}
\end{figure*}
\end{widetext}
\begin{figure*}
	\includegraphics[scale=0.5]{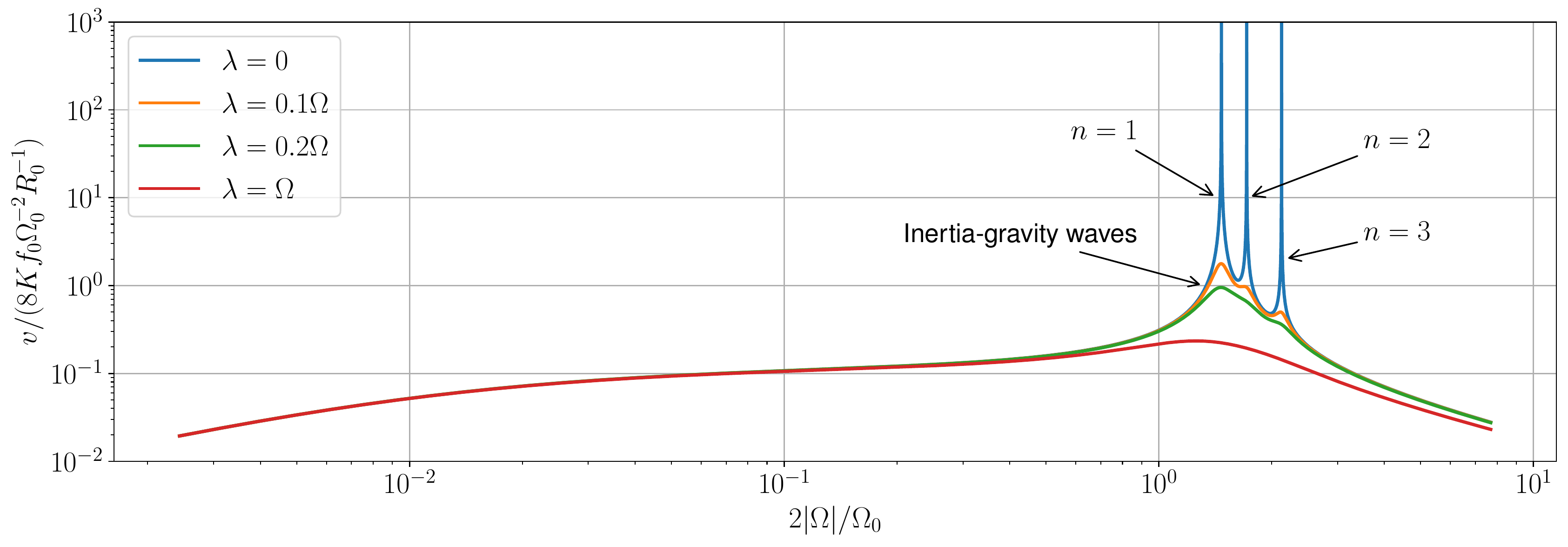}
	\includegraphics[scale=0.5]{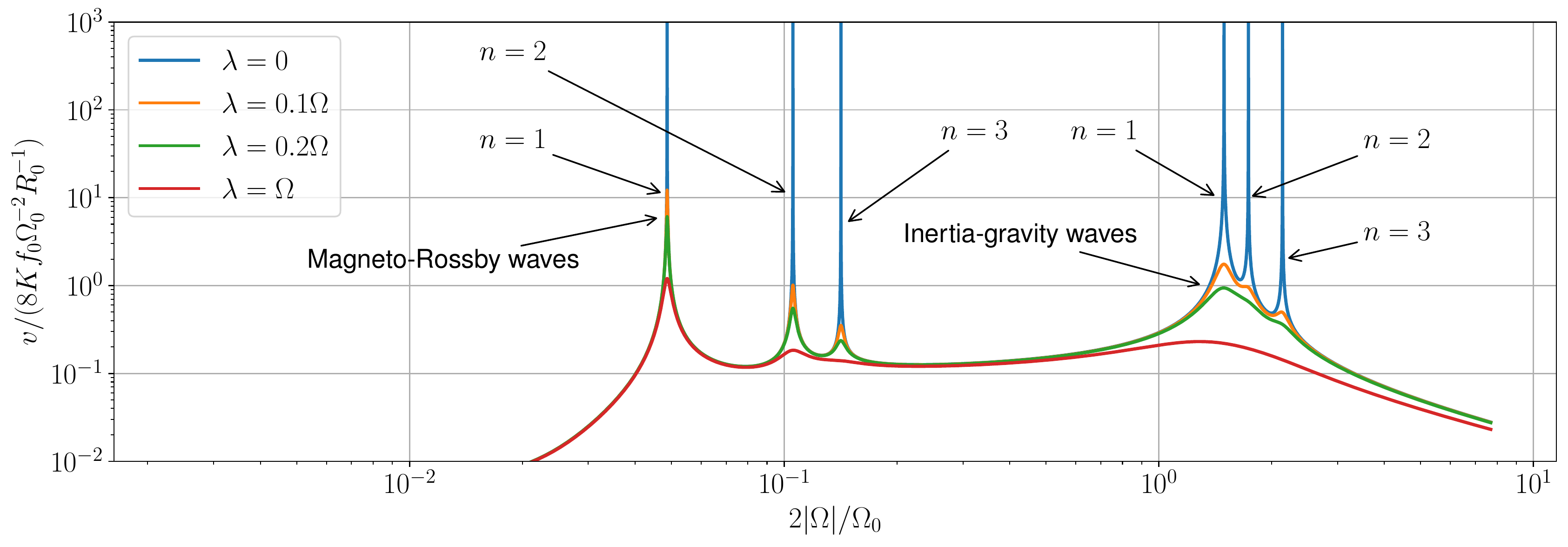}
	\caption{Normalized velocity amplitudes due to solution (\ref{eq:SolMid}) for $v_A = 0\, {\rm cm}\, {\rm s}^{-1}$ (top) and $v_A = 12.6\cdot 10^3 \, {\rm cm}\, {\rm s}^{-1}$ (bottom) as a function of the normalized forcing frequency $\Omega$. The calculated frequency range includes the first three prograde propagating magneto-Rossby and magneto-inertia wave modes. We used  $\Omega_0 = 26\cdot 10^{-7}\, {\rm s}^{-1}$, $R_0 = 5\cdot 10^{10}\, {\rm cm}$, $C_0 = 13 \cdot 10^{3}\, {\rm cm}\, {\rm s}^{-1}$, $\beta = 1.04 \cdot 10^{-16}\, {\rm cm}^{-1}\, {\rm s}^{-1}$.}
	\label{fig:ForcedPro}
\end{figure*}
The introduced coefficients $\bar{\Lambda}_1, \bar{\Lambda}_2, \bar{\Lambda}_3, \bar{K}_1, \bar{K}_2$ are specified in Appendix \ref{sec:SolMid}. In the solution, the symmetric part $\sim \cos(...)$ describes non-resonant waves between the resonances, and the antisymmetric part $\sim \sin(...)$ captures resonant waves in the vicinity of the eigenfrequencies, respectively. As a remarkable difference to the equatorial solution (\ref{eq:SolEq}), the responses (\ref{eq:SolMid}) do not fade out in the limit of zero Coriolis forces $f_0 = \beta = 0$ and zero gravity $C_0 =0$, showing that Alfv\'{e}n waves can be directly excited by tidal forces. In the limit of pure Alfv\'{e}n waves $\Lambda_n$ simplifies to,
\begin{align}
	\Lambda_n = \frac{8i(-1)^{n-1}K\Omega}{R_0 (2n-1)\pi\left[4\Omega^2 + 2i\lambda \Omega - \frac{4v_A^2}{R_0^2}\right]}, \nonumber 
\end{align} 
which yields, by reintroducing the series (\ref{eq:FSeries}),  the rather simple, unidirectional wave solution 
\begin{align}
	v = \frac{2K\Omega}{R_0 \left[4\Omega^2 + 2i\lambda \Omega - \frac{4v_A^2}{R_0^2}\right]}\exp\left(i\frac{2x}{R_0} - i2\Omega t\right). \label{eq:SolAlfv}
\end{align} 
This solution is independent of the tachocline layer thickness $H_0$ and does further not rely on the presence of any reduced gravity $g$ such that Equation (\ref{eq:SolAlfv}) is not confined to our shallow water model and can be used in a more universal way to estimate tidally excited Alfv\'{e}n waves in stars, regardless of the exact strata properties. At resonances Alfv\'{e}n waves have maximum amplitudes of $v = K/\lambda$ so that further modeling efforts to estimate the magnetic and viscous dissipation of Alfv\'{e}n waves $\lambda_{\text{Alfv\'{e}n}}$ (that is physically different to planetary waves) seems necessary for future studies.
\begin{figure*}
	\includegraphics[scale=0.52]{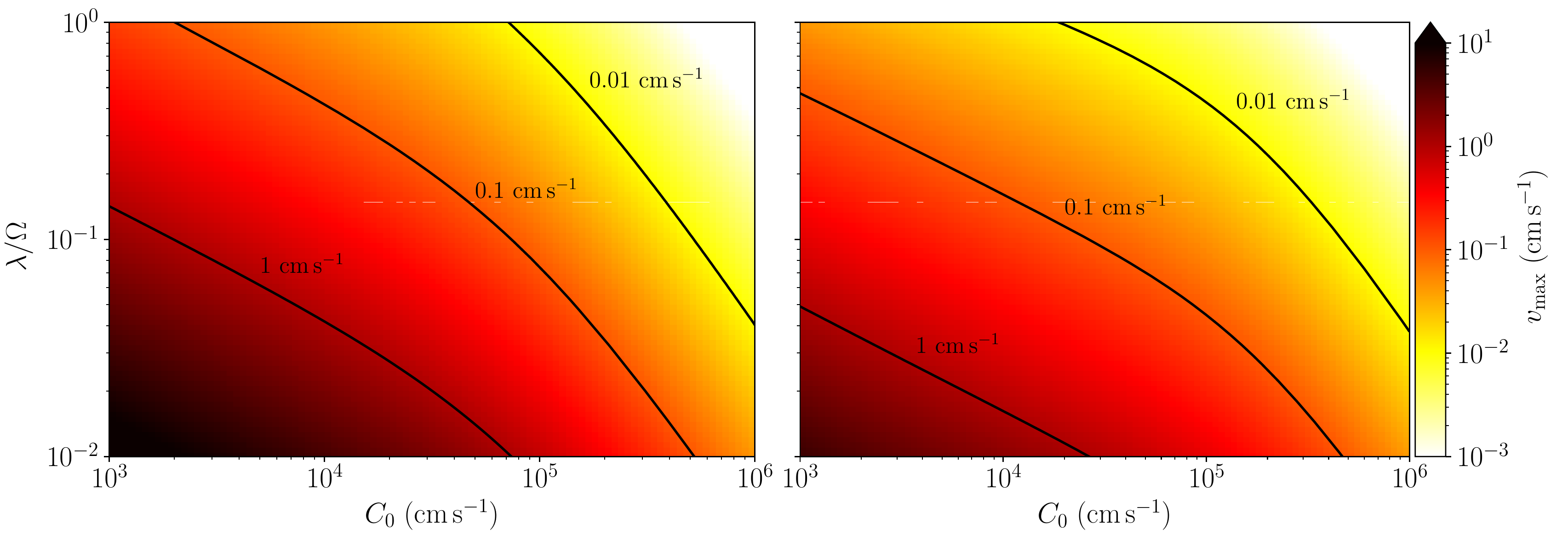}
	\caption{Resonant velocity amplitudes of the first $n=1$ retrograde wave mode for Rossby (left) and inertia-gravity (right) waves as a function of $C_0$ and $\lambda$. The parameters $v_A = 0\, {\rm cm}\, {\rm s}^{-1}$, $\Omega_0 = 26\cdot 10^{-7}\, {\rm s}^{-1}$, $R_0 = 5\cdot 10^{10}\, {\rm cm}$, $\beta = 1.04 \cdot 10^{-16}\, {\rm cm}^{-1}\, {\rm s}^{-1}$ and $K = 504\,{\rm cm}^{2}\, {\rm s}^{-2}$ were used for the calculation.}
	\label{fig:ResponseEq}
\end{figure*}

Similarly as in section \ref{sec:ForcedEq}, we study the wave responses graphically through calculating normalized amplitudes as a function of the tidal frequency $\Omega$ for the first three wave modes $n=1,2,3$ and for different damping ratios. Since, in contrast to equatorial waves, the retrograde solution branch comprising classic Rossby waves and the prograde branch describing magneto-Rossby waves differ significantly, we study retrograde and prograde waves separately. Retrograde wave responses are shown in Figure \ref{fig:ForcedRetro} without (top) and with (bottom) the presence of a toroidal magnetic field. As the most striking difference to equatorial waves, it appears that the first $n=1$ Rossby mode resonates with an orders of magnitude higher amplitude than any other wave mode, which has made it necessary to present the velocity amplitudes on a logarithmic scale. Higher mode responses and non-resonant waves seem to be insignificant at mid latitudes. When a magnetic field is applied, the Rossby eigenfrequencies are shifted towards higher frequencies and, rather interestingly, we find an inverted dispersion, meaning that the eigenfrequencies increase with growing wave numbers instead of being reduced. In Figure \ref{fig:ForcedPro} we also show the prograde wave responses in the hydrodynamic case (top) and under the effect of the toroidal magnetic field (bottom). Prograde Rossby waves do not exist in the hydrodynamic limit, which is why only high-frequency responses of inertia-gravity waves remain visible. In the magnetic case, we find almost the same response pattern as for retrograde waves, with the small difference that the eigenfrequencies are slightly smaller. We can conclude that both progradely and retrogradely excited Rossby waves may resonate to a similar extent\textemdash an intriguing difference to analogous geophysical wave responses, where the tidal sense always plays a pivotal role. 
\section{Estimation of wave responses in our sun} \label{sec:Sun}
In the previous sections we have analyzed characteristic tidal wave excitations for the general class of solar-like stars. The amplitudes were normalized to keep the analysis as general as possible, but we had to use fixed values for the parameters $C_0$ and $v_A$ to calculate the response patterns, which both affect the eigenfrequencies and thereby shift the resonance peaks. Apart from this effect, the presented wave responses are generally valid and we can easily deduce, e.g., that the first latitudinal wave mode $n=1$ is invariably showing the most significant response. That was certainly to be expected due the large-scale nature of the tidal potential, but the difference to the second mode $n=2$ is substantial. In order to detect planetary waves in solar-like stars which are mainly excited by some tidally dominant planet, e.g., (hot) Jupiters, our analysis suggests to search for $m,n = 2,1$ responses, for which the tidal energy input is by far the highest in the leading order. Starting from this insight, we want to estimate the maximum attainable velocity amplitudes in our Sun, which can result from tidal forcing. The tidal forces experienced by the Sun are mainly dictated by a complex interplay of Jupiter, Venus, Earth (and Mercury), which involves many different excitation frequencies, among them the solar rotation, the individual orbit frequencies and, perhaps most intriguing, also tidal variations with periods of 11 years closely related to the solar cycle \citep{Okhlopkov2016}. At time-scales close to the solar rotation $\Omega_0$, waves are always excited retrogradely, whereas the tide-generating planets force planetary waves progradely at time scales in the order of their orbit periods.
Here, we simplify these complex dynamics by considering Jupiter as the sole tide-generating body (exactly the case for which we derived the tidal potentials (\ref{eq:PotNon}) and (\ref{eq:PotEq})) and taking the forcing frequency $\Omega$ as arbitrary in the first instance. The latter idealization is justified in view of the argument that the buoyancy frequency (governed by the effective gravity) increases from zero to several rotations per minute as we descend in the tachocline from the convection zone down to the radiative interior. This means there is always a place in the tachocline where the buoyancy period can match the tidal period, or, to put it another way, the tachocline can in principal resonate to any given excitation frequency. In the following we calculate resonant amplitudes as a function of the unspecified quantities $C_0$ and $\lambda$, for which we first calculate the eigenfrequencies $\omega_{n=1}(C_0 , \lambda)$ using Equations (\ref{eq:DispEqDamp}) and (\ref{eq:DispDampFull}) and inserting them into the solutions (\ref{eq:SolEq}) and (\ref{eq:SolMid}).
\begin{figure*}
	\includegraphics[scale=0.52]{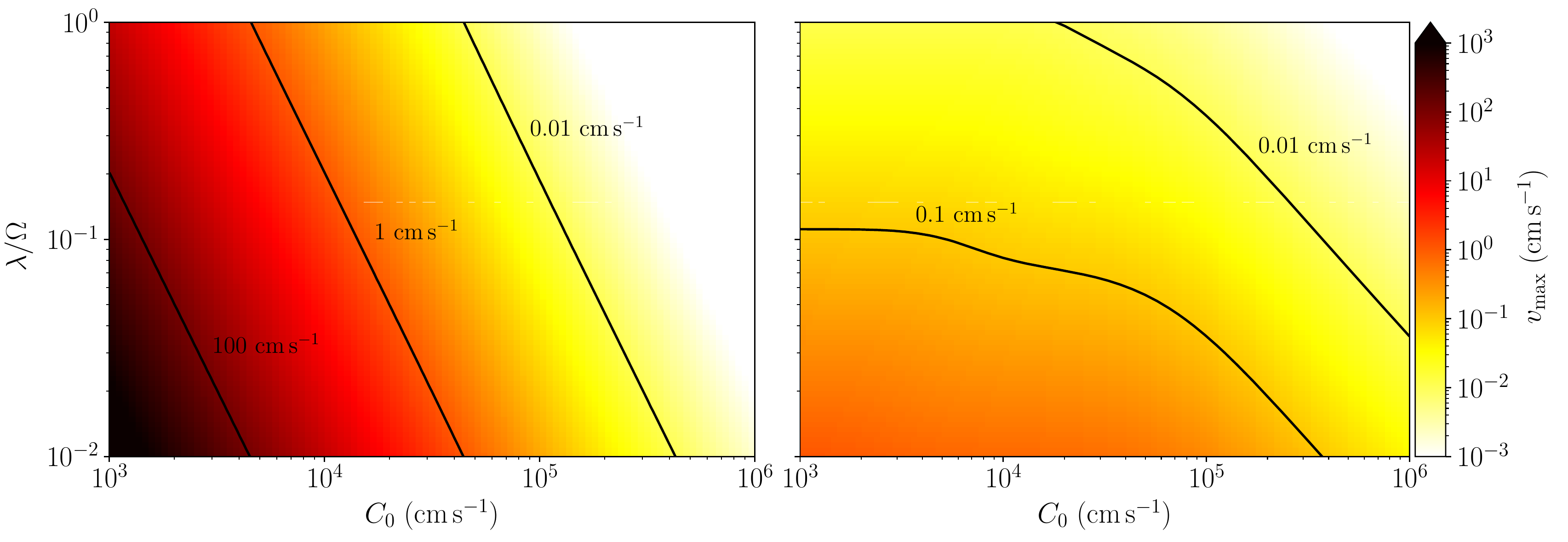}
	\caption{Resonant velocity amplitudes of the first $n=1$ retrograde wave mode for Rossby (left) and inertia-gravity (right) waves as a function of $C_0$ and $\lambda$. The parameters $v_A = 0\, {\rm cm}\, {\rm s}^{-1}$, $\Omega_0 = 26\cdot 10^{-7}\, {\rm s}^{-1}$, $R_0 = 5\cdot 10^{10}\, {\rm cm}$, $\beta = 1.04 \cdot 10^{-16}\, {\rm cm}^{-1}\, {\rm s}^{-1}$ and $K = 504\,{\rm cm}^{2}\, {\rm s}^{-2}$ were used for the calculation.}
	\label{fig:ResponseMid}
\end{figure*}
\subsection{Equatorial waves}
We start to analyze resonant equatorial wave excitations. At first it needs to be noted that non-resonant wave responses, as they occur in the large frequency range between Rossby and gravity inertia wave, see Figure \ref{fig:Forced}, are about $v \approx 0.02\, {\rm cm}\, {\rm s}^{-1}$. Such low values are entirely negligible for the solar dynamics, so that we will indeed have to focus on resonant excitations. Resonant velocity amplitudes for the full range of gravity velocities considered in the literature and damping ratios between $0.01$ and $1$ are shown in Figure \ref{fig:ResponseEq} for both hydrodynamic Rossby (left) and gravity-inertia (right) waves. 
We find a large range of possible velocity amplitudes $10^{-3}\, {\rm cm}\, {\rm s}^{-1} \lesssim v \lesssim 10\, {\rm cm}\, {\rm s}^{-1}$, where in particular low-frequency Rossby waves in the regime $10^{3}\, {\rm cm}\, {\rm s}^{-1} \lesssim C_0 \lesssim 10^4\, {\rm cm}\, {\rm s}^{-1}$, which correspond to periods being in the order of the Schwabe cycle, can reach amplitude velocities above $v = 1\, {\rm cm}\, {\rm s}^{-1}$. Such velocities may already be dynamo effective, however, all in all, tidal excitations of equatorial waves seem to play only a minor role compared to other mechanism capable of stimulating planetary waves. The available helioseismic data allows the extraction of $m=2$ Rossby waves only for RMS velocities larger $v > 50\, {\rm cm}\, {\rm s}^{-1}$ \citep{Liang2019}, such that tidally forced waves at the equator are unverifiable.
Up to now, only sectional modes $m=n$ in the range $3 \leq m \leq 15$ have been detected \citep{Loptien2018}.

Prograde magnetic Rossby waves are largely suppressed by constant toroidal fields resulting in even weaker wave response not deserving any further discussion. At this point it seems very promising for future studies to incorporate nonuniform latitudinal magnetic field profiles into the model as done by \cite{Zaqarashvili2018a}, allowing to study very slow prograde magneto-Rossby waves, which can reach Schwabe-cycle periods for many different combinations of gravity and Alfv\'{e}n velocities.
\begin{figure*}
	\includegraphics[scale=0.52]{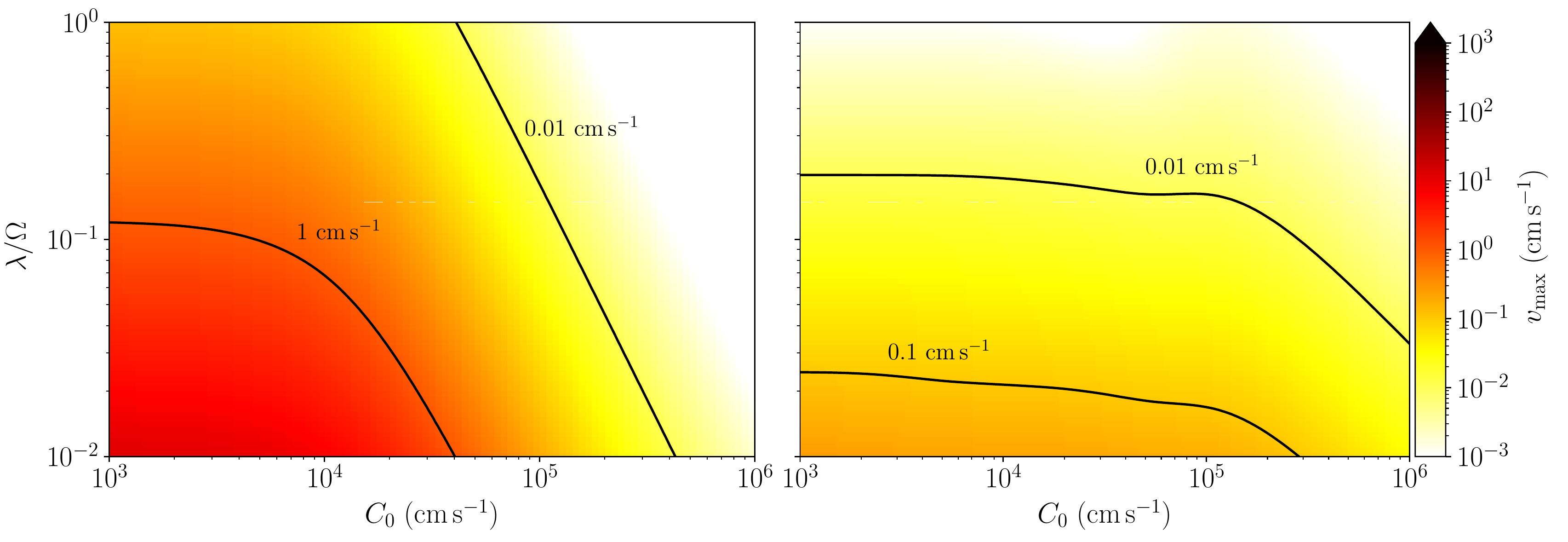}
	\caption{Resonant velocity amplitudes of the first $n=1$ prograde wave mode for magneto Rossby (left) and magneto inertia-gravity (right) waves as a function of $C_0$ and $\lambda$. The parameters $v_A = 12.6\cdot 10^3\, {\rm cm}\, {\rm s}^{-1}$, $\Omega_0 = 26\cdot 10^{-7}\, {\rm s}^{-1}$, $R_0 = 5\cdot 10^{10}\, {\rm cm}$, $\beta = 1.04 \cdot 10^{-16}\, {\rm cm}^{-1}\, {\rm s}^{-1}$ and $K = 504\,{\rm cm}^{2}\, {\rm s}^{-2}$ were used for the calculation.}
	\label{fig:ResponseMidMHD}
\end{figure*}
\subsection{Non-equatorial waves}
At mid latitudes, we find still smaller non-resonant velocities about $v \approx 0.004\, {\rm cm}\, {\rm s}^{-1}$, which about five times smaller than the corresponding non-resonant equatorial velocities. However, this finding does not allow us to draw any conclusions about general response tendencies. As already discussed in section \ref{sec:MidResponse}, $n=1$ Rossby waves can reach extraordinary high amplitudes at resonance despite the low non-resonant amplitude level. Indeed, Figure \ref{fig:ResponseMid} confirms that Rossby waves can (theoretically) resonate with amplitudes of more than $v = 1\, {\rm m}\, {\rm s}^{-1}$, whereas gravity-inertia waves are even less excited than their equatorial counterparts.
In this order of magnitude tidally forced Rossby waves would indeed be capable of providing sufficient energy to considerably affect the solar dynamics and then to take part in the synchronization process of the dynamo, especially since the tidal energy input of the other essential planets Venus and Earth have been disregarded here.
\begin{figure*}
	\includegraphics[scale=0.52]{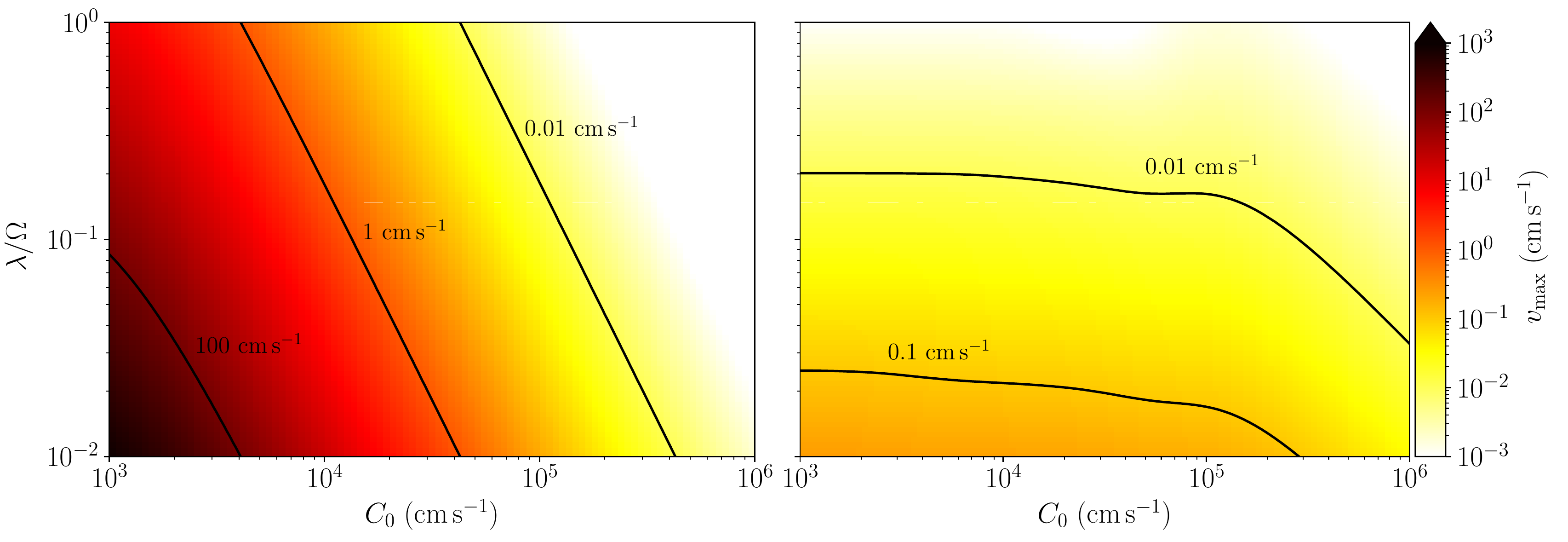}
	\caption{Resonant velocity amplitudes of the first $n=1$ prograde wave mode for magneto Rossby (left) and magneto inertia-gravity (right) waves as a function of $C_0$ and $\lambda$. The parameters $v_A = 1\cdot 10^3\, {\rm cm}\, {\rm s}^{-1}$, $\Omega_0 = 26\cdot 10^{-7}\, {\rm s}^{-1}$, $R_0 = 5\cdot 10^{10}\, {\rm cm}$, $\beta = 1.04 \cdot 10^{-16}\, {\rm cm}^{-1}\, {\rm s}^{-1}$ and $K = 504\,{\rm cm}^{2}\, {\rm s}^{-2}$ were used for the calculation.}
	\label{fig:ResponseMidMHDSlow}
\end{figure*}

At mid latitudes, also prograde magnetic waves can respond significantly to tidal excitations. For the strong magnetic field scenarios $B_0 = 20\, {\rm kG} \equiv v_A = 12.6\cdot 10^3\, {\rm cm}\, {\rm s}^{-1}$ discussed before, we find nevertheless considerably lower velocity amplitudes up to $v = 10\, {\rm cm}\, {\rm s}^{-1}$, see Figure \ref{fig:ResponseMidMHD}, which is comparable to the weak equatorial wave responses. However, very slow magneto Rossby waves with periods in the order of the solar cycle and longer are impacted by sufficiently lower toroidal magnetic fields of $B_0 \sim 1 {\rm kG}$, as they can be generated by steady non-reversing dynamos or be manifested as an offshoot of the primordial field in the radiative interior, which may penetrate into the tachocline \citep{Zaqarashvili2015}. Therefore, in Figure \ref{fig:ResponseMidMHDSlow} we also show the resonant responses of prograde waves subject to a slow Alfv\'{e}n velocity of $v_A = 1\cdot 10^3\, {\rm cm}\, {\rm s}^{-1}$. In this scenario, the amplitudes are comparable to their retrograde counterparts and may be dynamo-effective in a larger range of the $\lambda /\Omega$\textendash$C_0$ space.  
\subsection{Plausibility of the responses}\label{sec:Plau}
We have discussed possible wave responses in the solar tachocline under different scenarios for a wide range of the unspecified quantities $\lambda$ and $C_0$. The anticipated amplitudes vary in each case over several orders of magnitude, so that up to this point it is not possible to make a valid statement about the specific tidal energy input; only excitation potentials have been demonstrated. The presented velocity maps also do not allow drawing any probability information as to which damping regimes can be reached. We shall therefore briefly discuss the plausibility of the presented scenarios. First, it is important to be aware that the velocity amplitudes are always smaller for small $C_0$ primarily due to the fact that small $C_0$ are associated with small eigenfrequencies, meaning that the absolute damping rates $\lambda$ also become smaller since we have considered constant damping ratios $\lambda/ \Omega$. This poses the question of whether this assumption is realistic, whether low-frequency and high-frequency waves actually have comparable lifetimes. The answer depends on the nature of damping. Although solar fluid dynamics is of strongly turbulent nature, leading-order damping of large-scale Rossby waves is likely governed by laminar viscous boundary layers. In turbulent convection, the kinematic viscosity should be understood as an eddy viscosity \citep{Ruediger1989}, whose exact value is controversial in the literature. \cite{Gizon2020} have applied the value $\nu \approx 250\, {\rm km}^{2}\, {\rm s}^{-1}$ associated with supergranular scales to account for the shearing boundary layers arising from differential rotation. For turbulent viscosities of such high orders, boundary layers are expected to be essentially laminar and we can think of oscillatory Stokes-like boundary layers forming on top of the rigid radiative interior, in which the horizontal velocities fall off exponentially to fulfill the no-slip boundary condition. When damping is dominated by such oscillatory layers, the damping rate scales as $\lambda \sim \sqrt{\Omega}$, such that low frequency waves (corresponding to small $C_0$) are indeed subject to weaker damping. However, the quantities $\lambda/\Omega$ and $C_0$ are only independent if $\lambda \sim \Omega$, which leads us to conclude that $\lambda /\Omega$ will be distinctly higher for small $C_0$ than for large gravity velocities. Indeed, \cite{Liang2019} found, in average, longer lifetimes for high-frequency waves at lower Rossby modes than for low-frequency waves at higher modes. Nevertheless, the overall damping behavior remains largely unsettled, all the more so since thermal and magnetic dissipation may also make significant contributions. Therefore, it seems expedient for future studies to explicitly calculate the different Stokes and Ekman boundary layers, similar to the authors \cite{Bildsten2000} for the case of neutron stars, in order to provide better estimates of viscous damping rates, which, apart from the calculation of resonant wave responses, could also allow for better modeling of horizontal eigenfunctions observed at the solar surface \citep{Proxauf2020}. 

Finally, we need to take a closer look at the direction of excitation. We presented response scenarios for both prograde and retrograde planetary waves, however, when focusing on the tidally dominant planet Jupiter, we only have two frequencies $\Omega_0$ and $\Omega_t$ involved. Obviously, the sun's angular frequency is orders of magnitude higher than Jupiter's orbit frequency $\Omega_0 \gg \Omega_t$, letting us approximate $\Omega = \Omega_t - \Omega_0 \approx - \Omega_0$. Apparently only the retrograde branches, comprising classic Rossby and gravity-inertial waves, will be stimulated directly. The most relevant and promising scenario to reach significantly high forced velocity amplitudes is therefore represented in Figure \ref{fig:ResponseMid} in the form of hydrodynamic waves excited at mid latitudes. However, when considering the  tidal forces of Jupiter, Venus and Earth in conjunction, far lower forcing frequencies appear in envelope curves of the potentials, most relevantly the 11 year period closely matching to the solar activity periodicity \citep{Okhlopkov2016}.  Envelope variations of the combined tidal potentials can also excite planetary waves progradely in the most susceptible low $C_0$ regimes\textemdash best requirements for the excitation of slow magneto-Rossby waves. At this point it seems worthwhile to incorporate the tidal forcing of all three planets and to calculate multifrequential wave responses numerically. We plan to accomplish this in a future study. As a second interesting scenario of low-frequency forcing, we can also think off solar-like stars forced by tidally-synchronized planets, or likewise binary stars, where the dominant frequency $\Omega_0$ is absent and slow frequencies, both prograde and retrograde, may remain as a result of orbital changes. For all these systems it is conceivable that low-frequency tidal forcing may dictate the activity cycles, be it by synchronizing the dynamo or by directly implanting large-scale atmospheric motions, provided that the forcing amplitude $K$ is large enough (resp. the damping rate $\lambda$ small enough) to induce sufficiently high wave amplitudes.

\section{Concluding remarks}
By equipping the magnetohydrodynamic ``shallow water'' equations with a tidal potential term and linear friction, we have constructed a first theoretical set-up for the study of damped and forced planetary waves in the tachocline layer of solar-like stars. The governing equations were projected onto two different Cartesian planes in the vicinity of the equator and at mid latitudes, allowing us to describe both equatorially trapped and locally unbounded waves in the most approachable way. As a key result, we have shown that the linearized system of governing equations can be combined into one decoupled wave equation (\ref{eq:Wave}) for the local latitudinal velocity component, which markedly simplifies the Fourier analysis for extracting the characteristic dispersion relations in different wave limits of interest.

We solve this wave equation analytically within different regimes, starting with the known free wave solutions via damped wave dynamics up to the complete forced wave problem. The analysis revealed that the damping behavior of magnetohydrodynamic planetary waves is more intricate than the damping of classical geophysical waves since the introduced damping parameter can translate into very different decay rates for the different wave types. Most interesting here is our finding that the damping rates of retrograde Rossby waves at mid-latitudes correspond exactly to the decay rates, whereas progradely propagating magnetic Rossby waves are predicted to decay considerably slower by a factor of the squared natural eigenfrequency divided by the squared Alfv\'{e}n frequency. Further, the damped wave solutions derived for equatorial waves facilitate to calculate the meridional scales as a function of the damping rate and the toroidal magnetic field, with the result that damping always widens and the presence of magnetic fields always, in an even overcompensating way, narrows the equatorial waveguide.

The forced wave problem is solved analytically for the idealized case of a single tide-generating body prescribing a perfect circular orbit around the central star. The solutions can describe both non-resonant and resonant wave responses, the latter, however, are largely determined by the \textit{a priori} unknown damping coefficient. We found that for fixed damping ratios equatorial waves always respond with higher velocity amplitudes than mid-latitudes waves under non-resonant conditions, whereas mid-latitude waves have higher peak velocities in proximity to resonances. Among all types of planetary waves, the first large-scale Rossby mode, be it the classic retrograde or magnetic prograde Rossby wave, is found to always resonate with the highest amplitudes when considering fixed lifetimes. Rossby waves are therefore confirmed to be indeed a most promising candidate to potentially act as a resonance ground for low-frequency tidal excitations. 

Finally, we applied the solutions to the specific scenario of our Sun tidally forced by Jupiter for estimating possible velocity responses. We obtained  non-resonant amplitudes of $v \approx 0.02\, {\rm cm}\, {\rm s}^{-1}$ at the equator and of $v \approx 0.004\, {\rm cm}\, {\rm s}^{-1}$ at mid-latitudes, which are, for the solar dynamo, completely negligible. Resonant amplitudes strongly depend on both damping and the effective gravity (the considered region in the tachocline layer) so that resonant velocities deviate by several orders of magnitude with respect to these parameters. Consequently, a reliable prediction of the anticipated responses cannot yet be made without further ado. However, we found that for low-frequency excitations in the order of the 11 year solar cycle, with particular view on the 11.07-years alignment periodicity of the tidally dominant planets Venus, Earth,
and Jupiter, the tidal energy input of Jupiter alone can evoke high Rossby wave amplitudes of $v \gtrsim 100\, {\rm cm}\, {\rm s}^{-1}$ for fairly small damping ratios in the range $0.01 \lesssim \lambda /\Omega \lesssim 0.1$. We can conclude that, despite the fact that tidal accelerations are very small, significant velocities can potentially be induced through resonant amplification if dissipation is sufficiently small.  

To draw more definite conclusions, our model must be extended in different directions. Our pilot analysis has stressed the potential that tidal forcing may induce significant tachoclinic wave motions, but to irrevocably verify or disprove this possibility  further modeling of the different dissipation sources, including viscous, turbulent, thermal and magnetic damping, is required. In addition, it is expedient to solve the problem in spherical coordinates and to include the full potentials of all significant planets, which would allow to determine the exact response to alignment periodicities and spring tides. Regardless of the exact attainable amplitude levels, there also remains the question of how and in what way Rossby waves may affect the solar dynamo. The linear Rossby wave solutions do not produce net kinetic helicity, which is the main ingredient for the $\alpha$ effect. Hence, it appears instructive to look at the nonlinear evolution of tidally excited Rossby waves, e.g., tachocline nonlinear oscillations (TNOs) arising from the energy exchange between Rossby waves and differential rotation \citep{Dikpati2018}. Moreover, it might be promising to estimate how wavelike displacements of the tachocline layer affect the entropy stratification to check if tide stimulated flux tube instabilities can possibly encroach into the dynamo process \textcolor{red}{\citep{Ferriz-Mas1996,Charbonneau2022}}. Finally, coupling the planetary wave equations with dynamo models may provide valuable insights into possible synchronization mechanisms imparted by Rossby waves.

\begin{acknowledgments} 
This work has received funding from the European Research
Council (ERC) under the European Union’s Horizon 2020 research and innovation
programme (grant agreement No 787544). It was also supported in frame of the Helmholtz - RSF Joint Research Group “Magnetohydrodynamic instabilities: Crucial relevance for large scale liquid metal batteries and the sun-climate connection”, contract No HRSF-0044. TVZ was supported by the Austrian Fonds zur F{\"o}rderung der Wissenschaftlichen Forschung (FWF) project P30695-N27 and by Shota Rustaveli
National Science Foundation of Georgia (project FR-21-467).
\end{acknowledgments}

%

\vspace{5mm}
\facilities{}


\software{}



\newpage
\appendix

\section{Derivation of the wave equation} \label{sec:WaveDer}
Differentiation of equations (\ref{eq:TD1}) and (\ref{eq:TD2}) with respect to time and using equations (\ref{eq:TD3}) and (\ref{eq:TD4}) yields the coupled differential equations for the velocities
\begin{align}
	\frac{\partial^2 u}{\partial t^2} - f\frac{\partial v}{\partial t} = v_A^2 \frac{\partial^2 u}{\partial x^2} + C_0^2 \left(\frac{\partial^2 u}{\partial x^2} + \frac{\partial^2 v}{\partial x \partial y}\right) -\frac{\partial^2 V}{\partial x \partial t} - \lambda \frac{\partial u}{\partial t}, \\
	\frac{\partial^2 v}{\partial t^2} + f\frac{\partial u}{\partial t} = v_A^2 \frac{\partial^2 v}{\partial x^2} + C_0^2 \left(\frac{\partial^2 u}{\partial x \partial y} + \frac{\partial^2 v}{\partial^2 y}\right) -\frac{\partial^2 V}{\partial y \partial t} - \lambda \frac{\partial v}{\partial t}, \label{eq:A2}
\end{align}
to be decoupled in the following. For this we can replace the term $f\partial_t u$ in (\ref{eq:A2}) by equation (\ref{eq:TD1}), giving
\begin{align}
\square_{v_A} v +f^2 v-fg\frac{\partial \eta}{\partial x} + f \frac{B_0}{4\pi \rho}\frac{\partial b_x}{\partial x} - f\frac{\partial V}{\partial x} - f\lambda u = C_0^2 \frac{\partial}{\partial y}\left(\frac{\partial u}{\partial x} + \frac{\partial v}{\partial y}\right) -\frac{\partial^2 V}{\partial y \partial t} - \lambda \frac{\partial v}{\partial t}. \label{eq:A3}
\end{align}
As an intermediate step, we calculate $\partial_y (\ref{eq:TD1})$ - $\partial_x (\ref{eq:TD2})$ to extract the following expression
\begin{align}
	\frac{\partial \eta}{\partial t} = \frac{H_0}{f}\frac{\partial \zeta}{\partial t} + \lambda\frac{H_0}{f} \zeta + \frac{H_0}{f} \beta v - \frac{H_0}{f}\frac{B_0}{4\pi \rho} \frac{\partial}{\partial x}\left(\frac{\partial b_y}{\partial x} - \frac{\partial b_x}{\partial y} \right), \label{eq:A4}
\end{align}
where $\zeta = \partial_x v - \partial_y u$ denotes the horizontal vorticity. We can now differentiate (\ref{eq:A3}) with respect to time and insert (\ref{eq:A4}) to eliminate $\eta$
\begin{align}
	\frac{\partial}{\partial t}\square_{v_A} v +f^2 \frac{\partial v}{\partial t}- C_0^2 \frac{\partial^2 \zeta}{\partial t \partial x} - C_0^2 \lambda \frac{\partial \zeta}{\partial x} - C_0^2 \beta \frac{\partial V}{\partial x} + C_0^2\frac{B_0}{4\pi \rho} \frac{\partial^2}{\partial x^2}\left(\frac{\partial b_y}{\partial x} - \frac{\partial b_x}{\partial y} \right) + f v_A^2 \frac{\partial^2 u}{\partial x^2} - f\frac{\partial^2 V}{\partial t\partial x} - f\lambda \frac{\partial u}{\partial t} \nonumber \\
	 = C_0^2 \frac{\partial}{\partial t}\frac{\partial}{\partial y}\left(\frac{\partial u}{\partial x} + \frac{\partial v}{\partial y}\right) -\frac{\partial^3 V}{\partial y \partial t^2} - \lambda \frac{\partial^2 v}{\partial t^2}. \label{eq:A5}
\end{align} 
By noting that 
\begin{align}
	C_0^2\frac{\partial^2 \zeta}{\partial t \partial x} + C_0^2 \frac{\partial}{\partial t}\frac{\partial}{\partial y}\left(\frac{\partial u}{\partial x} + \frac{\partial v}{\partial y}\right) = C_0^2 \frac{\partial}{\partial t}\Delta v
\end{align}
and taking again the time derivative, we find 
\begin{align}
	\frac{\partial^2}{\partial t^2}\square_{v_A} v +f^2 \frac{\partial^2 v}{\partial t^2}- C_0^2 \frac{\partial^2}{\partial t^2}\Delta v - C_0^2 \lambda \frac{\partial^2 \zeta}{\partial t\partial x} - C_0^2 \beta \frac{\partial^2 V}{\partial t \partial x} + C_0^2 v_A^2 \left(\frac{\partial^4 v}{\partial x^4} - \frac{\partial^4 u}{\partial x^3 \partial y} \right) + f v_A^2 \frac{\partial^3 u}{\partial t\partial x^2}\nonumber \\
	 - f\frac{\partial^3 V}{\partial t^2 \partial x} - f\lambda \frac{\partial^2 u}{\partial t^2} 
	  +\frac{\partial^4 V}{\partial y \partial t^3} + \lambda \frac{\partial^3 v}{\partial t^3} = 0. \label{eq:A7}
\end{align}
Let us collect all remaining $u$ velocity terms in (\ref{eq:A7}), reading
\begin{align}
	C_0^2 \lambda \frac{\partial^3 u}{\partial x \partial y \partial t} - C_0^2 v_A^2 \frac{\partial^4 u}{\partial x^3 \partial y} + f v_A^2 \frac{\partial^3 u}{\partial t\partial x^2} - f\lambda \frac{\partial^2 u}{\partial t^2} = \left(v_A^2 \frac{\partial^2}{\partial x^2} - \lambda \frac{\partial}{\partial t} \right)\left(f \frac{\partial u}{\partial t} - C_0^2\frac{\partial^2 u}{\partial x \partial y}\right).
\end{align}
Thankfully, the two terms in the right bracket can be eliminated simply by rearranging equation (\ref{eq:A2}) in the following form
\begin{align}
	f \frac{\partial u}{\partial t} - C_0^2\frac{\partial^2 u}{\partial x \partial y} = v_A^2 \frac{\partial^2 v}{\partial x^2} - \frac{\partial^2 v}{\partial t^2} + C_0^2\frac{\partial^2 v}{\partial^2 y} -\frac{\partial^2 V}{\partial y \partial t} - \lambda \frac{\partial v}{\partial t}. \label{eq:A9}
\end{align}
This way, we have finally eliminated all $u$ velocity terms. Inserting Equation (\ref{eq:A9}) into Equation (\ref{eq:A7}) yields the wave equation (\ref{eq:Wave}) presented in section \ref{sec:Formulation}.

\section{Derivation of the tidal potential} \label{sec:Tidal}
\begin{figure}
	\begin{center}
		\includegraphics[width=0.75\textwidth]{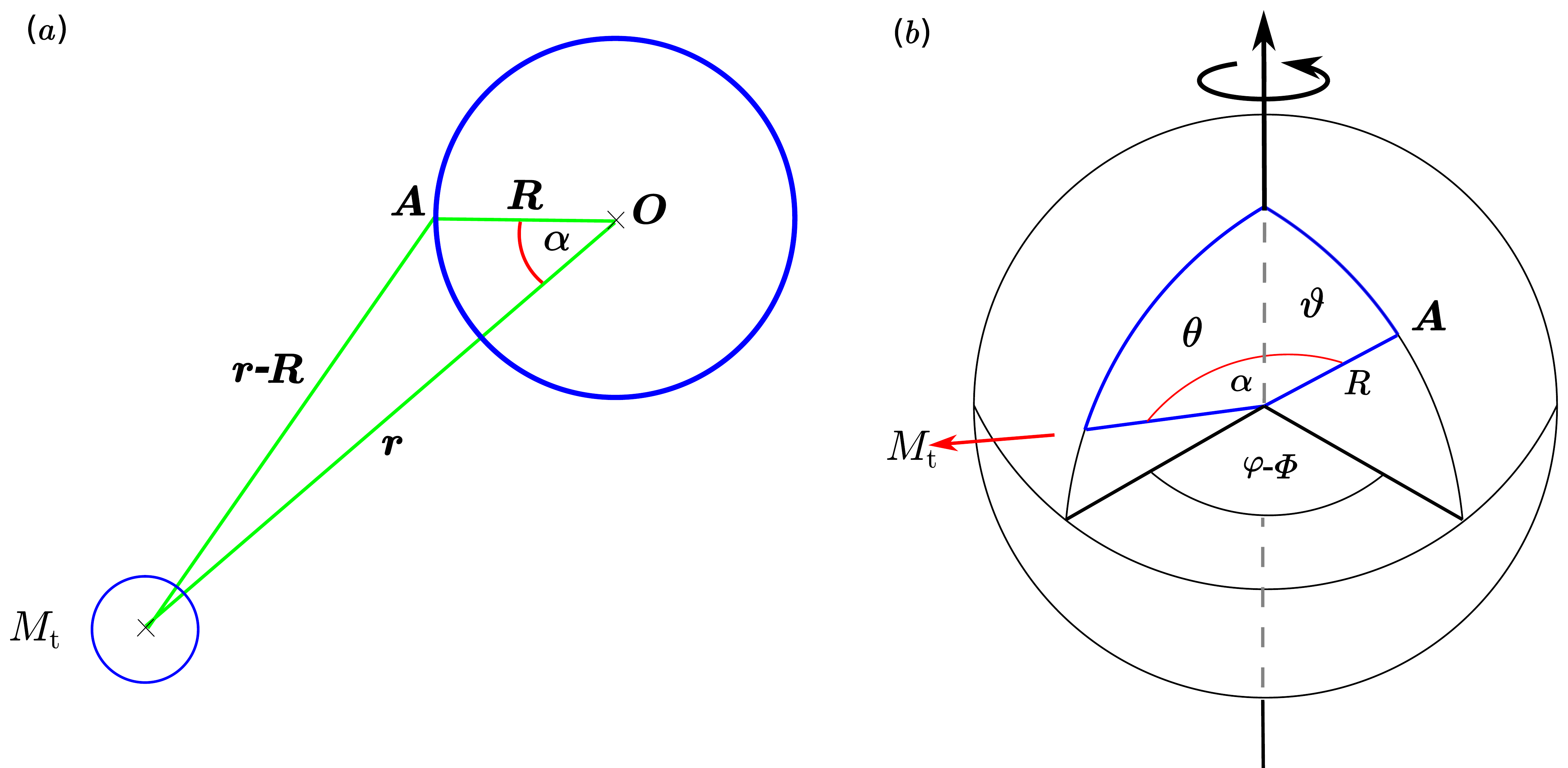}
		\caption{\label{fig::sketch_tidal_pot}
			(a) Sketch of a two body system with definition of quantities that
			are involved in the calculation of the tidal potential caused by $M_{\rm{t}}$ on the
			surface of a central body. (b) Definition of angles in a spherical
			system of reference.
		}
	\end{center}
\end{figure}

We consider a central body with radius $R_0$ with its center of mass
being the origin labeled $\vec{O}$ and a tide
generating body with mass $M_{\rm{t}}$ at a distance $r=r(t)$ from the
$\vec{O}$, see Figure \ref{fig::sketch_tidal_pot}a. The tidal force originating from $M_{\rm{t}}$
results from the combined action of the pseudoforce of inertia due to
the {\it{free fall}} motion of the central body around the center of
mass of the system and the gravitational force of the perturbing body.
At location $\vec{A}$ on the surface of the central object the tidal
force can be computed from the vector difference of the gravitational
pull the perturbing body exerts on a test object at position $\vec{A}$
and the gravitational pull the perturbing body would exert on this
object at the center of the central body (which corresponds to
the pseudo force of inertia of the free fall motion).
The tidal force is a conservative force and can be expressed in terms
of a scalar potential, $\vec{F}_{\rm{t}}=-\nabla V_{\rm{t}}$, which
can be developped in terms of an infinite sum of Legendre polynomials
labeled by their degree $l$. Since the terms with $l=0$ and $l=1$ do not
contribute to the force and the relation of the radius $R_0$ to the
distance $r$ is small, only one term with $l=2$ has to be
considered (see, e.g., \cite{agnew2015}). Then the tidal potential is
\begin{equation}
	V_{\rm{t}}(A)=\frac{G
		M_{\rm{t}}R_0^2}{r^3(t)}P_2(\cos\alpha),
	\label{eq::tidal_pot}
\end{equation}
where $G$ is the gravitational constant and $P_2$ is the
Legendre polynom with degree $l=2$. The angle $\alpha$ represents the
solid angle between the line connecting the origin $\vec{O}$ and the
observation position $\vec{A}$ and the line connecting $\vec{O}$ and
the center of the tide generating object
(Figure \ref{fig::sketch_tidal_pot}b).  
It is convenient to transfer the problem into
a spherical coordinate system, where we have two sets of variables:
$\varphi$ and $\theta$, which denote the longitude and the latitude
coordinate of the observation point $\vec{A}$, and $\varTheta$ and
$\varPhi$, which describe the location of the tide generating object
$M_{\rm{t}}$.  The Legendre polynom
$P_l(\cos\alpha)$ can then be expressed as a sum of associated
Legendre polynomials in $\vartheta,\varphi,\varTheta$ and $\varPhi$
via the expansion in terms of spherical harmonics: 
\begin{equation}
	P_l(\cos\alpha)=
	\frac{1}{2l+1}\sum\limits_{m=-l}^{l}
	\overline{Y_l^{m}}(\vartheta,\varphi)Y_l^m(\varTheta,\varPhi)
	\mbox{   with   }
	Y_l^m(\vartheta,\varphi)=(-1)^m\left[\frac{2l+1}{4\pi}\frac{(l-m)!}{(l+m)!}\right]^{{1}/{2}}
	P_l^m(\cos\vartheta)e^{im\varphi},
\end{equation}
where $P_l^m$ are the associated Legendre functions taken from \cite{munk1966}.
Since we only perform a decomposition for $l=2$, we only need the
following five associated Legendre polynomials
\begin{equation}
	P_2^{-2}  =  \frac{1}{8}\sin^2\vartheta,
	\qquad
	P_2^{-1}  =  \frac{1}{2}\sin\vartheta\cos\vartheta,
	\qquad
	P_2^{0}  =  \frac{1}{2}(3\cos^2\vartheta-1),
	\qquad
	P_2^{1}  =  -3\sin\vartheta\cos\vartheta,
	\qquad
	P_2^{2}   =  3\sin^2\vartheta ,
\end{equation}
so that after some calculations using standard relations for the trigonometric
functions we obtain
\begin{equation}
	P_2(\cos\alpha)  =
	\frac{3}{4}\left[\frac{1}{3}(3\cos^2\varTheta-1)(3\cos^2\vartheta-1)
	+\sin2\varTheta\sin2\vartheta\cos(\varphi-\varPhi)
	+\sin^2\varTheta\sin^2\vartheta\cos(2\varphi-2\varPhi)\right].
	\label{eq::legendre2}
\end{equation}
We now consider a tide generating object that moves exactly in the
plane defined by the equatorial plane of the central body so that
$\varTheta=90^{\circ}$. Thus the second term in the brackets on the
right side of Eq.~(\ref{eq::legendre2}) ($\propto \sin2\varTheta$)
vanishes and the first term simplyfies to $\sin^2\vartheta$ (after
dropping the constant terms that do not contribute to the tidal force.  We further assume that the distance between central
body and perturbing body is constant, i.e. $r(t)=r$.  In reality the
two bodies propagate around a common center of mass on a path that is
the solution of the corresponding Kepler problem.  The resulting
variation of the distance between both bodies introduces a slow time
scale for the amplitude of the tidal force, which can be neglected
with regard to the current problem. Hence, the tidal potential finally 
reads:
\begin{equation}\label{eq::tidpot}
	V_{\rm{t}}(R,\vartheta,\varphi,t) = 
	\underbrace{\frac{3}{4}\frac{GM_{\rm{t}}}{R}\left(\frac{R}{r}\right)^3}_{\displaystyle
		K} \left[\sin^2\vartheta\left[1+\cos\left(2\varphi-2\varPhi\right)\right]\right],
\end{equation}
where we have introduced the constant term $K$. The only time-dependent
quantity in Equation (\ref{eq::tidpot}) is the angle $\varPhi$, which
describes the longitude of the tide generating body with respect to
the origin.  The time-dependence of $\varPhi$ is simply given by
$\varPhi(t)=\left(\varOmega_{\rm{t}}-\varOmega_{\rm{0}}\right)t$ where
$\varOmega_{\rm{0}}$ denotes the angular frequency of the rotation of
the central body (i.e. the sun in our particular case) and
$\Omega_{\rm{t}}$ the angular frequency of the motion of the
perturbing body around the central object.
In the following we translate the angular coordinates $\varphi$ and
$\vartheta$ into the Cartesian coordinates $x$ and $y$ as they appear
in the $\beta$-plane model.  We consider a small regime around a fixed
latitudinal coordinate $\vartheta_0$ so that we can write
$\vartheta=\vartheta_0+\vartheta'$ with $\vartheta'=y/R_0 \ll 1$.
The corresponding expansions of the
expression that describes the latitudinal dependence is then
\begin{equation}
	\sin^2(\vartheta)=  \sin^2(\vartheta_0+\vartheta')
	\approx \sin^2(\vartheta_0)+\sin(2\vartheta_0)\frac{y}{R_0}
	+\cos(2\vartheta_0)\frac{y^2}{R_0^2}+...
\end{equation}
In order to keep the same approximation used for the $\beta$-plane
approach we ignore the term quadratic in $y$.  We examine two
particular cases. At the equator we have $\vartheta_0=\pi/2$ so that
$\sin^2(\vartheta_0)=1$ and $\sin(2\vartheta_0)=0$ and we end up with
a forcing independent of $y$.
\begin{equation}
	V_{\rm{t}}^{\rm{eq}}(x,y,t) = 
	K\left[1+\cos\left[2x-2\left(\varOmega_{\rm{t}}-\varOmega_{\rm{0}}\right)t\right]\right] .
\end{equation}
For non-equatorial solutions it is reasonable to choose
$\vartheta_0=\pi/4$ so that $\sin^2(\vartheta_0)=1/2$ and
$\sin(2\vartheta_0)=1$ and the corresponding tidal potential reads
\begin{equation}
	V_{\rm{t}}^{45^{\circ}}(x,y,t) = 
	K \left(\frac{1}{2}+\frac{y}{R_0}\right)
	\left[1+\cos\left[2x-2\left(\varOmega_{\rm{t}}-\varOmega_{\rm{0}}\right)t\right]\right].
\end{equation}
\newpage

\section{Forced wave solutions at the equator}
\label{sec:SolEq}
The forced wave problem is solved by inserting the ansatz
\begin{equation}
	v = \sum_{m=0}^{\infty}\sum_{n=0}^{\infty} \alpha_{m,n}(t)\exp\left(\frac{im}{R_0}x\right)\exp\left(-\frac{\mu y^2}{2}\right)H_n (\sqrt{\mu}y), \label{eq:AnsatzA}
\end{equation} 
and the coordinate expansion 
\begin{equation}
	y = \sum_{n=1}^{\infty}\frac{2^{\frac{5}{2}-2n}}{\sqrt{\mu}(n-1)!}\exp\left(-\frac{\mu y^2}{2}\right)H_{2n-1}(\sqrt{\mu}y). \label{eq:SeriesA}
\end{equation}
into the wave equation (\ref{eq:WaveEq}). The comparison of the modal coefficients $\alpha_{m,n}(t)$ with the forcing potential $\sim \cos(2x/R_0 - 2\Omega t)$ directly reveals the zonal wave number to be fixed at $m = 2/R_0$, which describes planetary waves with two crests and troughs fitting around the equator. All other modal coefficients are not subject to any forcing and consequently obey damped wave equations, letting them decay exponentially in time. Linear stationary solutions can only be found for $m = 2$. In the same way we can argue that latitudinal wave numbers corresponding to stationary solutions must be uneven $n\rightarrow 2n+1$ since only uneven Hermitian polynomials appear in the expansion (\ref{eq:SeriesA}). The remaining modal coefficients must satisfy the conditional equation
\begin{align}
	&\sum_{n=1}^{\infty} \ddddot{\alpha}_{2,2n-1}(t)\frac{4\Omega^2}{C_0^2}\exp\left(-\frac{\mu y^2}{2}\right)H_{2n-1}(\sqrt{\mu}y)\exp\left(i\frac{2x}{R_0}\right) \nonumber \\
	+&\sum_{n=1}^{\infty} \dddot{\alpha}_{2,2n-1}(t)\frac{8 \lambda\Omega^2}{C_0^2}\exp\left(-\frac{\mu y^2}{2}\right)H_{2n-1}(\sqrt{\mu}y)\exp\left(i\frac{2x}{R_0}\right) \nonumber \\
	+&\sum_{n=1}^{\infty} \ddot{\alpha}_{2,2n-1}(t)\frac{32v_A^2\Omega^2}{C_0^2 R_0^2}\exp\left(-\frac{\mu y^2}{2}\right)H_{2n-1}(\sqrt{\mu}y)\exp\left(i\frac{2x}{R_0}\right) \nonumber \\
	+&\sum_{n=1}^{\infty} \alpha_{2,2n-1}(t)\frac{64v_A^2\Omega^2}{R_0^4}\exp\left(-\frac{\mu y^2}{2}\right)H_{2n-1}(\sqrt{\mu}y)\exp\left(i\frac{2x}{R_0}\right) \nonumber \\
	+&\sum_{n=1}^{\infty} \alpha_{2,2n-1}(t)\frac{64v_A^4\Omega^2}{C_0^2 R_0^4}\exp\left(-\frac{\mu y^2}{2}\right)H_{2n-1}(\sqrt{\mu}y)\exp\left(i\frac{2x}{R_0}\right) \nonumber \\
	+&\sum_{n=1}^{\infty} \ddot{\alpha}_{2,2n-1}(t)\frac{16\Omega^2}{R_0^2}\exp\left(-\frac{\mu y^2}{2}\right)H_{2n-1}(\sqrt{\mu}y)\exp\left(i\frac{2x}{R_0}\right) \nonumber \\
	-&\sum_{n=1}^{\infty} \ddot{\alpha}_{2,2n-1}(t)\left[4\Omega^2 +2\lambda \Omega i-\frac{4v_A^2}{R_0^2}\right]\textcolor{blue}{\frac{\partial^2}{\partial y^2}}\exp\left(-\frac{\mu y^2}{2}\right)H_{2n-1}(\sqrt{\mu}y)\exp\left(i\frac{2x}{R_0}\right) \nonumber \\
	+&\sum_{n=1}^{\infty} \ddot{\alpha}_{2,2n-1}(t)\textcolor{blue}{\frac{4\beta^2 \Omega^2}{C_0^2} y^2}\exp\left(-\frac{\mu y^2}{2}\right)H_{2n-1}(\sqrt{\mu}y)\exp\left(i\frac{2x}{R_0}\right) \nonumber \\
	+&\sum_{n=1}^{\infty} \ddot{\alpha}_{2,2n-1}(t)\frac{32\lambda \Omega^2 v_A^2}{C_0^2 R_0^2}\exp\left(-\frac{\mu y^2}{2}\right)H_{2n-1}(\sqrt{\mu}y)\exp\left(i\frac{2x}{R_0}\right) \nonumber \\
	+&\sum_{n=1}^{\infty} \dot{\alpha}_{2,2n-1}(t)\frac{16\lambda \Omega^2}{ R_0^2}\exp\left(-\frac{\mu y^2}{2}\right)H_{2n-1}(\sqrt{\mu}y)\exp\left(i\frac{2x}{R_0}\right) \nonumber \\
	+&\sum_{n=1}^{\infty} \ddot{\alpha}_{2,2n-1}(t)\frac{4\lambda^2 \Omega^2}{ C_0^2}\exp\left(-\frac{\mu y^2}{2}\right)H_{2n-1}(\sqrt{\mu}y)\exp\left(i\frac{2x}{R_0}\right) \nonumber \\
	+&\sum_{n=1}^{\infty} \dot{\alpha}_{2,2n-1}(t)\frac{8i\beta\Omega^2}{R_0}\exp\left(-\frac{\mu y^2}{2}\right)H_{2n-1}(\sqrt{\mu}y)\exp\left(i\frac{2x}{R_0}\right) \nonumber \\
	=&\sum_{n=1}^{\infty}\frac{32K\beta \Omega^4}{C_0^2 R}\frac{2^{\frac{5}{2}-2n}}{\sqrt{\mu}(n-1)!}\exp\left(-\frac{\mu y^2}{2}\right)H_{2n-1}(\sqrt{\mu}y)\exp\left(i\frac{2x}{R_0} - i2\Omega t\right). \label{eq:Modal}
\end{align}
We have used $\alpha_{2,2n-1} = i\dot{\alpha}_{2,2n-1}/2\Omega$ and $\alpha_{2,2n-1} = -\ddot{\alpha}_{2,2n-1}/4\Omega^2$ to evaluate the three terms involving $y$ derivatives.
From equation (\ref{eq:Modal}) we want to extract ordinary differential equations for the modal coefficients $\alpha_{2,2n-1}(t)$. All terms are proportional to $\exp(i2x/R_0)$ so that the $x$ coordinate drops out by default. This is unfortunately not the case for the $y$ coordinate since the blue-marked terms do not fit with the Hermitian base. However, we can bypass this issue by applying the identity
\begin{align}
	\mu^2 y^2 \exp\left(-\frac{\mu y^2}{2}\right)H_n(\sqrt{\mu}y) - \frac{\partial^2}{\partial y^2}\exp\left(-\frac{\mu y^2}{2}\right)H_n(\sqrt{\mu}y) = (2n+1)\mu \exp\left(-\frac{\mu y^2}{2}\right)H_n(\sqrt{\mu}y) \label{eq:Identity}
\end{align}
that projects the two terms on the Hermitian eigenbasis. Inserting $\mu$ (\ref{eq:Mu}) allows to rearrange (\ref{eq:Identity}) into
\begin{align}
	\left[\frac{4\Omega^2 + 2i\lambda\Omega \beta^2}{C_0^2}y^2 - \left(4\Omega^2 - \frac{4v_A^2}{R_0^2}\right)\frac{\partial^2}{\partial y^2}\right]\exp\left(-\frac{\mu y^2}{2}\right)H_{2n-1}(\sqrt{\mu}y) \nonumber \\ = \left[\frac{2\Omega \beta}{C_0}\sqrt{4\Omega^2 + 2i\lambda \Omega - \frac{4v_A^2}{R_0^2}}(4n-1)\right]\exp\left(-\frac{\mu y^2}{2}\right)H_{2n-1}(\sqrt{\mu}y) \label{eq:Transform}
\end{align}    
we insert into (\ref{eq:Modal}) to replace the blue marked terms, yielding 
\begin{align}
	\sum_{n=1}^{\infty}\Bigg[\ddddot{\alpha}_{2,2n-1}(t) + 2\lambda \dddot{\alpha}_{2,2n-1}(t) + \left(\frac{C_0 \beta}{2\Omega}\sqrt{4\Omega^2 +2i\lambda \Omega - \frac{4v_A^2}{R_0^2}}(4n-1) + \frac{4}{R_0^2}(C_0^2 + 2v_A^2) + \frac{8\lambda v_A^2}{R_0^2} + \lambda^2 \right)\ddot{\alpha}_{2,2n-1}(t)\nonumber \\
	+\left(\frac{4\lambda C_0^2}{R_0^2} - \frac{2iC_0^2 \beta}{R_0} \right)\dot{\alpha}_{2,2n-1}(t) + \left(\frac{16 v_A^2 C_0^2}{R_0^4} + \frac{16 v_A^4}{R_0^4}\right)\alpha_{2,2n-1}(t) \nonumber \\
	-\frac{8K\beta \Omega^2}{ R}\frac{2^{\frac{5}{2}-2n}}{\sqrt{\mu}(n-1)!}\exp(-i2\Omega t)\Bigg]\exp\left(-\frac{\mu y^2}{2}\right)H_{2n-1}(\sqrt{\mu}y) = 0
\end{align}
The infinite sums can only yield zero if all individual summands disappear. Therefore,
each coefficient $\alpha_{2,2n-1}(t)$ must satisfy the following set of ordinary differential equations
\begin{align}
	\ddddot{\alpha}_{2,2n-1}(t) + 2\lambda \dddot{\alpha}_{2,2n-1}(t) + \left(\frac{C_0 \beta}{2\Omega}\sqrt{4\Omega^2 +2i\lambda \Omega - \frac{4v_A^2}{R_0^2}}(4n-1) + \frac{4}{R_0^2}(C_0^2 + 2v_A^2) + \frac{8\lambda v_A^2}{R_0^2} + \lambda^2 \right)\ddot{\alpha}_{2,2n-1}(t)\nonumber \\
	+\left(\frac{4\lambda C_0^2}{R_0^2} - \frac{2iC_0^2 \beta}{R_0} \right)\dot{\alpha}_{2,2n-1}(t) + \left(\frac{16 v_A^2 C_0^2}{R_0^4} + \frac{16 v_A^4}{R_0^4}\right)\alpha_{2,2n-1}(t) 
	-\frac{8K\beta \Omega^2}{ R}\frac{2^{\frac{5}{2}-2n}}{\sqrt{\mu}(n-1)!}\exp(-i2\Omega t) = 0 \label{eq:ModalA}
\end{align}
We have now reduced the partial differential Equation (\ref{eq:WaveEq}) into an infinite set of decoupled ordinary differential equations to be solved readily. Equation (\ref{eq:ModalA}) has the following stationary solution:
\begin{align}
	&\alpha_{2,2n-1}(t) = \frac{8iK\beta \Omega^2 2^{\frac{5}{2}-2n}(R\sqrt{\mu}(n-1)!)^{-1}\exp(-i2\Omega t)}{\left[ \left(4\Omega^2 +2i\lambda \Omega - \frac{4 v_A^2}{R_0^2}\right)\left(4\Omega^2 + 2i\lambda \Omega -\frac{4}{R_0^2}(C_0^2 + v_A^2)\right) - C_0^2\frac{4\beta}{R_0}\Omega - C_0 \beta|2\Omega|\sqrt{4\Omega^2 +2i\lambda \omega - \frac{4v_A^2}{R_0^2}}(4n-1)\right]}.
\end{align}
Inserting the modal coefficients back into the ansatz (\ref{eq:AnsatzA}) yields the solution (\ref{eq:SolEq}) we present in section \ref{sec:ForcedEq}. 
\section{Forced wave solutions at mid latitudes}\label{sec:SolNonEq}
We solve the forced wave problem by inserting the ansatz
\begin{equation}
	v = \sum_{m=0}^{\infty}\sum_{n=0}^{\infty}\left[\alpha_{m,n}(t)\sin\left(\frac{m}{R_0}x\right) + \beta_{m,n}(t)\cos\left(\frac{m}{R_0}x\right) \right] \exp\left(\frac{in\pi}{R_0}y\right) \label{eq:AnsatzB}
\end{equation} 
to be confined within a mid-latitude wave channel defined by the interval $-R_0 /2 \leq y \leq R_0 /2$, into the wave equation  (\ref{eq:WaveNonEq}). The forcing terms on the right-hand side of (\ref{eq:WaveNonEq}) involve both constant and $y$-proportional terms, which both must be expanded as Fourier series (\ref{eq:FSeries}) and (\ref{eq:FSeries2}) in order to project them onto the harmonic basis used in the ansatz.  
The direct comparison of the modal coefficients $\alpha_{m,n}(t)$ and $\beta_{m,n}(t)$ with the forcing potential $\sim \cos(2x/R_0 - 2\Omega t)$ directly reveals the zonal wave number to be fixed at $m = 2/R_0$, which describes planetary waves with two crests and troughs fitting around the equator. All other modal coefficients are not subject to any forcing and consequently obey freely damped wave equations, letting them decay exponentially in time. Linear stationary solutions can only be found for $m = 2$. In the same way we can argue that latitudinal wave numbers corresponding to stationary solutions must be uneven $n\rightarrow 2n+1$ since only uneven terms appear in the Fourier expansion (\ref{eq:FSeries}) and (\ref{eq:FSeries2}). The remaining modal coefficients must satisfy the conditional equation
\begin{align}
	&\sum_{n=1}^{\infty}\left[\ddddot{\alpha}_{2,2n-1}(t)\sin\left(\frac{2x}{R_0}\right) + \ddddot{\beta}_{2,2n-1}(t)\cos\left(\frac{2x}{R_0}\right) \right] \exp\left(\frac{i(2n-1)\pi}{R_0}y\right) \nonumber \\
	+&\sum_{n=1}^{\infty} \left[\dddot{\alpha}_{2,2n-1}(t)\sin\left(\frac{2x}{R_0}\right) + \dddot{\beta}_{2,2n-1}(t)\cos\left(\frac{2x}{R_0}\right) \right]2\lambda\exp\left(\frac{i(2n-1)\pi}{R_0}y\right) \nonumber \\
	+&\sum_{n=1}^{\infty} \left[\ddot{\alpha}_{2,2n-1}(t)\sin\left(\frac{2x}{R_0}\right) + \ddot{\beta}_{2,2n-1}(t)\cos\left(\frac{2x}{R_0}\right) \right]\frac{8v_A^2}{R_0^2}\exp\left(\frac{i(2n-1)\pi}{R_0}y\right) \nonumber \\
	+&\sum_{n=1}^{\infty} \left[\ddot{\alpha}_{2,2n-1}(t)\sin\left(\frac{2x}{R_0}\right) + \ddot{\beta}_{2,2n-1}(t)\cos\left(\frac{2x}{R_0}\right) \right]\frac{4C_0^2}{R_0^2}\exp\left(\frac{i(2n-1)\pi}{R_0}y\right) \nonumber \\
	+&\sum_{n=1}^{\infty} \left[\ddot{\alpha}_{2,2n-1}(t)\sin\left(\frac{2x}{R_0}\right) + \ddot{\beta}_{2,2n-1}(t)\cos\left(\frac{2x}{R_0}\right) \right]\frac{C_0^2(2n-1)^2\pi^2}{R_0^2}\exp\left(\frac{i(2n-1)\pi}{R_0}y\right) \nonumber \\
	+&\sum_{n=1}^{\infty} \left[\ddot{\alpha}_{2,2n-1}(t)\sin\left(\frac{2x}{R_0}\right) + \ddot{\beta}_{2,2n-1}(t)\cos\left(\frac{2x}{R_0}\right) \right]f_0^2\exp\left(\frac{i(2n-1)\pi}{R_0}y\right) \nonumber \\
	+&\sum_{n=1}^{\infty} \left[\ddot{\alpha}_{2,2n-1}(t)\sin\left(\frac{2x}{R_0}\right) + \ddot{\beta}_{2,2n-1}(t)\cos\left(\frac{2x}{R_0}\right) \right]\lambda^2\exp\left(\frac{i(2n-1)\pi}{R_0}y\right) \nonumber \\
	+&\sum_{n=1}^{\infty} \left[ \dot{\beta}_{2,2n-1}(t)\sin\left(\frac{2x}{R_0}\right) - \dot{\alpha}_{2,2n-1}(t)\cos\left(\frac{2x}{R_0}\right) \right]\frac{2C_0^2 \beta}{R_0}\exp\left(\frac{i(2n-1)\pi}{R_0}y\right) \nonumber \\
	+&\sum_{n=1}^{\infty} \left[\dot{\alpha}_{2,2n-1}(t)\sin\left(\frac{2x}{R_0}\right) + \dot{\beta}_{2,2n-1}(t)\cos\left(\frac{2x}{R_0}\right) \right]\frac{8\lambda v_A^2}{R_0^2}\exp\left(\frac{i(2n-1)\pi}{R_0}y\right) \nonumber \\
	+&\sum_{n=1}^{\infty} \left[\dot{\alpha}_{2,2n-1}(t)\sin\left(\frac{2x}{R_0}\right) + \dot{\beta}_{2,2n-1}(t)\cos\left(\frac{2x}{R_0}\right) \right]\frac{4\lambda C_0^2}{R_0^2}\exp\left(\frac{i(2n-1)\pi}{R_0}y\right) \nonumber \\
	+&\sum_{n=1}^{\infty} \left[\dot{\alpha}_{2,2n-1}(t)\sin\left(\frac{2x}{R_0}\right) + \dot{\beta}_{2,2n-1}(t)\cos\left(\frac{2x}{R_0}\right) \right]\frac{(2n-1)^2 \pi^2\lambda C_0^2}{R_0^2}\exp\left(\frac{i(2n-1)\pi}{R_0}y\right) \nonumber \\
	+&\sum_{n=1}^{\infty} \left[\alpha_{2,2n-1}(t)\sin\left(\frac{2x}{R_0}\right) + \beta_{2,2n-1}(t)\cos\left(\frac{2x}{R_0}\right) \right]\frac{16C_0^2v_A^2}{R_0^4}\exp\left(\frac{i(2n-1)\pi}{R_0}y\right) \nonumber \\
	+&\sum_{n=1}^{\infty} \left[\alpha_{2,2n-1}(t)\sin\left(\frac{2x}{R_0}\right) + \beta_{2,2n-1}(t)\cos\left(\frac{2x}{R_0}\right) \right]\frac{4(2n-1)^2 \pi^2 C_0^2v_A^2}{R_0^4}\exp\left(\frac{i(2n-1)\pi}{R_0}y\right) \nonumber \\
	+&\sum_{n=1}^{\infty} \left[\alpha_{2,2n-1}(t)\sin\left(\frac{2x}{R_0}\right) + \beta_{2,2n-1}(t)\cos\left(\frac{2x}{R_0}\right) \right]\frac{16v_A^4}{R_0^4}\exp\left(\frac{i(2n-1)\pi}{R_0}y\right) \nonumber \\
	=&\sum_{n=1}^{\infty}\left[\sin\left(\frac{2x}{R_0}\right)\cos\left(2\Omega t\right) - \cos\left(\frac{2x}{R_0}\right)\sin\left(2\Omega t\right) \right]\left(f_0 \Omega -\frac{2if_0 \Omega}{(2n-1)\pi} +2 \Omega^2 - \frac{2v_A^2}{R_0^2}\right)\frac{16K\Omega}{R_0}\frac{(-1)^{n-1}}{(2n-1)\pi}\exp\left(\frac{i(2n-1)\pi y}{R_0}\right)  \nonumber \\
	+&\sum_{n=1}^{\infty}\left[\sin\left(\frac{2x}{R_0}\right)\sin\left(2\Omega t\right) + \cos\left(\frac{2x}{R_0}\right)\cos\left(2\Omega t\right) \right] \frac{16K\lambda \Omega^2}{R_0}\frac{(-1)^{n-1}}{(2n-1)\pi}\exp\left(\frac{i(2n-1)\pi y}{R_0}\right)  \label{eq:Modal2}
\end{align}
All summands are either proportional to $\sim \sin(2x/R_0)\exp(i(2n-1)\pi y/R_0)$ or to $\sim \cos(2x/R_0)\exp(i(2n-1)\pi y/R_0)$ so that we can rearrange the terms as follows:
\begin{align}
	&\sum_{n=1}^{\infty}\Bigg[\ddddot{\alpha}_{2,2n-1}(t) + 2\lambda \dddot{\alpha}_{2,2n-1}(t) + \left(\frac{4C_0^2 + 8v_A^2 + (2n-1)^2 \pi^2 C_0^2}{R_0^2} + f_0^2 + \lambda^2\right)\ddot{\alpha}_{2,2n-1}(t)\nonumber \\
	+&\left(\frac{4\lambda C_0^2 + 8 \lambda v_A^2 + (2n-1)^2 \pi^2\lambda C_0^2}{R_0^2}\right)\dot{\alpha}_{2,2n-1}(t) + \frac{2\beta C_0^2}{R_0}\dot{\beta}_{2,2n-1}(t) + \left(\frac{16 v_A^2 C_0^2 + 16v_A^4 + 4(2n-1)^2 \pi^2 v_A^2 C_0^2 }{R_0^4}\right)\alpha_{2,2n-1}(t) \nonumber \\
	-&\left(f_0 \Omega -\frac{2if_0 \Omega}{(2n-1)\pi} +2 \Omega^2 - \frac{2v_A^2}{R_0^2}\right)\frac{16K\Omega}{R_0}\frac{(-1)^{n-1}}{(2n-1)\pi}\cos\left(2\Omega t\right) - \frac{16K\lambda \Omega^2}{R_0}\frac{(-1)^{n-1}}{(2n-1)\pi}\sin\left(2\Omega t\right) \Bigg] \nonumber \\
	 \times&\sin\left(\frac{2}{R_0}x\right)\exp\left(\frac{i(2n-1)\pi}{R_0}y\right) \nonumber \\
	 +&\sum_{n=1}^{\infty}\Bigg[\ddddot{\beta}_{2,2n-1}(t) + 2\lambda \dddot{\beta}_{2,2n-1}(t) + \left(\frac{4C_0^2 + 8v_A^2 + (2n-1)^2 \pi^2 C_0^2}{R_0^2} + f_0^2 + \lambda^2\right)\ddot{\beta}_{2,2n-1}(t)\nonumber \\
	 +&\left(\frac{4\lambda C_0^2 + 8 \lambda v_A^2 + (2n-1)^2 \pi^2\lambda C_0^2}{R_0^2}\right)\dot{\beta}_{2,2n-1}(t) - \frac{2\beta C_0^2}{R_0}\dot{\alpha}_{2,2n-1}(t) + \left(\frac{16 v_A^2 C_0^2 + 16v_A^4 + 4(2n-1)^2 \pi^2 v_A^2 C_0^2 }{R_0^4}\right)\alpha_{2,2n-1}(t) \nonumber \\
	 +&\left(f_0 \Omega -\frac{2if_0 \Omega}{(2n-1)\pi} +2 \Omega^2 - \frac{2v_A^2}{R_0^2}\right)\frac{16K\Omega}{R_0}\frac{(-1)^{n-1}}{(2n-1)\pi}\sin\left(2\Omega t\right) - \frac{16K\lambda \Omega^2}{R_0}\frac{(-1)^{n-1}}{(2n-1)\pi}\cos\left(2\Omega t\right) \Bigg] \nonumber \\
	 \times&\cos\left(\frac{2}{R_0}x\right)\exp\left(\frac{i(2n-1)\pi}{R_0}y\right) = 0
\end{align}
The infinite sums can only yield zero if all individual summands disappear independently. Therefore,
each pair of coefficients $\alpha_{2,2n-1}(t)$ and $\beta_{2,2n-1}(t)$ must satisfy the following set of coupled ordinary differential equations
\begin{align}
&\ddddot{\alpha}_{2,2n-1}(t) + 2\lambda \dddot{\alpha}_{2,2n-1}(t) + \tilde{\Lambda}_1\ddot{\alpha}_{2,2n-1}(t)
+\tilde{\Lambda}_2\dot{\alpha}_{2,2n-1}(t) + \frac{2\beta C_0^2}{R_0}\dot{\beta}_{2,2n-1}(t) + \tilde{\Lambda}_3\alpha_{2,2n-1}(t) \nonumber \\
&-\tilde{K}_1\cos\left(2\Omega t\right) - \tilde{K}_2\sin\left(2\Omega t\right) = 0, \label{eq:ModalAlpha}\\
&\ddddot{\beta}_{2,2n-1}(t) + 2\lambda \dddot{\beta}_{2,2n-1}(t) + \tilde{\Lambda}_1\ddot{\beta}_{2,2n-1}(t)
+\tilde{\Lambda}_2\dot{\beta}_{2,2n-1}(t) - \frac{2\beta C_0^2}{R_0}\dot{\alpha}_{2,2n-1}(t) + \tilde{\Lambda}_3\beta_{2,2n-1}(t) \nonumber \\
&+\tilde{K}_1\sin\left(2\Omega t\right) - \tilde{K}_2\cos\left(2\Omega t\right) = 0,\label{eq:ModalBeta}
\end{align}
where we have introduced the coefficients $\tilde{\Lambda}_1$, $\tilde{\Lambda}_2$, $\tilde{\Lambda}_3$, $\tilde{K}_1$ and $\tilde{K}_2$ for better readability, which are given as
\begin{align}
	&\tilde{\Lambda}_1 = \frac{4C_0^2 + 8v_A^2 + (2n-1)^2 \pi^2 C_0^2}{R_0^2} + f_0^2 + \lambda^2 , \nonumber \\
	&\tilde{\Lambda}_2 = \frac{4\lambda C_0^2 + 8 \lambda v_A^2 + (2n-1)^2 \pi^2\lambda C_0^2}{R_0^2}, \nonumber \\
	&\tilde{\Lambda}_3 = \frac{16 v_A^2 C_0^2 + 16v_A^4 + 4(2n-1)^2 \pi^2 v_A^2 C_0^2 }{R_0^4}, \nonumber \\
	&\tilde{K}_1 = \left(f_0 \Omega -\frac{2if_0 \Omega}{(2n-1)\pi} +2 \Omega^2 - \frac{2v_A^2}{R_0^2}\right)\frac{16K\Omega}{R_0}\frac{(-1)^{n-1}}{(2n-1)\pi}, \nonumber \ \ \ \tilde{K}_2 = \frac{16K\lambda \Omega^2}{R_0}\frac{(-1)^{n-1}}{(2n-1)\pi}.
\end{align}  
We have now reduced the partial differential Equation (\ref{eq:WaveNonEq}) into an infinite set of ordinary differential equations to be solved readily.
It is convenient to introduce auxiliary variables
\begin{equation}
a_1=\alpha_{2,2n-1}+i\beta_{2,2n-1}, ~~~ a_2=\alpha_{2,2n-1}-i\beta_{2,2n-1},	
\end{equation}
allowing to decouple the Equations (\ref{eq:ModalAlpha}) and (\ref{eq:ModalBeta}) into the following form:
\begin{align}
	&\ddddot{a_1}+2\lambda \dddot{a_1}+\bar{\Lambda}_1\ddot{a}_1+\left(\bar{\Lambda}_2  -\frac{2i\beta C_0^2}{R_0}\right)\dot{a}_1+\bar{\Lambda}_3a_1=(\bar{K}_1+i\bar{K}_2)\exp(-2i\Omega t), \\
	&\ddddot{a_2}+2\lambda \dddot{a_2}+\bar{\Lambda}_1\ddot{a}_2+\left(\bar{\Lambda}_2+\frac{2i\beta C_0^2}{R_0} \right)\dot{a}_2+\bar{\Lambda}_3a_2=(\bar{K}_1-i\bar{K}_2)\exp(2i\Omega t).
\end{align}
The particular, stationary solutions of these equations can be readily obtained 
\begin{align}
&a_1=\frac{\bar{K}_1+i\bar{K}_2}{16\Omega^4-4\Omega^2\bar{\Lambda}_1+\bar{\Lambda}_3-\frac{4\beta C_0^2\Omega}{R_0}+2i\Omega(8\lambda\Omega^2-\bar{\Lambda}_2)}\exp(-2i\Omega t)\\
&a_2=\frac{\bar{K}_1-i\bar{K}_2}{16\Omega^4-4\Omega^2\bar{\Lambda}_1+\bar{\Lambda}_3-\frac{4\beta C_0^2\Omega}{R_0}-2i\Omega(8\lambda\Omega^2-\bar{\Lambda}_2)}\exp(2i\Omega t).
\end{align}
As the last step, it remains to express the original coefficients $\alpha_{2,2n-1}$ and $\beta_{2,2n-1}$ as functions of $a_1$ and $a_2$ as
\begin{equation}
\alpha_{2,2n-1}=\frac{a_1+a_2}{2},~~~\beta_{2,2n-1}=\frac{a_1-a_2}{2i}.
\end{equation}
Inserting these modal coefficients back into the ansatz (\ref{eq:AnsatzB}) finally yields the solution (\ref{eq:SolMid}) we present in section \ref{sec:MidResponse}.
\label{sec:SolMid}

\bibliographystyle{aasjournal}



\end{document}